\documentclass[preprint,11pt]{aastex}
\usepackage{rotating}

\begin{document}
\title{Diffuse Cluster-Like Radio Emission in Poor Environments}
\author{Shea Brown \& Lawrence Rudnick}
\affil{Department of Astronomy, University of Minnesota, Minneapolis, MN  55455}

\begin{abstract} We present a study of the spectral, polarimetric, 
morphological and environmental properties of the diffuse radio source 
0809+39 using observations taken with the Westerbork Synthesis Radio 
Telescope, the Very Large Array, and archival optical and X-ray data. The 
source has two distinct diffuse, steep-spectrum components, one in the 
north that is highly polarized, and a linear southern component undetected 
in polarization.  We discuss several plausible origins for each component, 
and conclude that the northern bright polarized component is most likely a 
radio relic associated with a poor z$\sim$0.2 cluster of galaxies, with a 
radio/X-ray luminosity ratio two orders of magnitude above typical values. 
The southern component is aligned with a more extended filament of 
galaxies $\sim$5~Mpc long at z$\sim$0.04. Deep optical and X-ray 
follow-ups are still needed in order to confirm and understand the 
physical origins of the synchrotron emission. Whatever the details of 
these origins, 0809+39 highlights the utility of synchrotron radiation for 
illuminating the diffuse components of low density environments unrelated 
to rich clusters.  \end{abstract}


\clearpage

\section{Introduction} In recent years, large-scale diffuse radio sources 
have been discovered to be associated with the intracluster medium (ICM) 
of over 50 clusters of galaxies.  These features are unique probes of the 
non-thermal particle populations and magnetic fields within the cluster, 
and are believed to be important tracers of merger/formation dynamics. The 
majority of these sources were found by searching the environs of rich 
galaxy clusters for diffuse radio emission unassociated with active AGN 
\citep{giov99,kemp04}. But is there also a magnetized, relativistic plasma 
associated with lower density regions of the cosmic web, such as groups 
and filaments of galaxies? Cosmological simulations predict complicated 
networks of accretion/merger shocks in these low density regions, which 
could accelerate particles, compress magnetic fields, and illuminate 
magnetized plasma if it exists \citep[e.g.,][]{mini01,ryu03,pfro08}. 
Unbiased searches for synchrotron signatures of these shocks are needed to 
confirm these predictions.

One such diffuse source, of ambiguous origin, is 0809+39 \citep{dela06}, 
discovered through a blind search for diffuse radio emission in the WENSS 
survey \citep{rudn06}. The system showed no obvious associations with an 
active radio galaxy. Several poor clusters were found in the vicinity, 
which suggests a similar origin to radio relic and halo sources. However, 
lacking any evidence for cluster X-ray emission, its radio luminosity far 
exceeded the standard radio/x-ray ratios for these classes of sources 
\citep[e.g.,][]{giov04,rudn08}.

We present a systematic study of the spectral, polarimetric, morphological 
and environmental properties of 0809+39 using observations taken with the 
Westerbork Synthesis Radio Telescope (WSRT) and the Very Large Array 
(VLA). We explore the origin of the diffuse radio emission and evaluate 
where 0809+39 falls in the parameter space of known extragalactic radio 
sources. We discuss observations and data reduction in $\S$2, and in $\S$3 
we outline image production and analysis, including our application of 
Rotation Measure Synthesis \citep{bren05}. In $\S$4 we present archival 
optical and x-ray data, and in $\S$5 we discuss the implications of our 
findings, followed by a summary of our key messages.

For calculations in this paper, we assume $H_{o}=70$, 
$\Omega_{\Lambda}=0.7$, $\Omega_{M}=0.3$. We define the optically thin 
synchrotron spectrum as $F_{\nu} \propto \nu^{\alpha}$ throughout.
 
\section{Observations \& Data Reduction}

\subsection{Observations} The diffuse source 0809+39 was observed for 13 
hours over two nights in P-band (350~MHz) with the Westerbork Synthesis 
Radio Telescope (WSRT) in January of 2006. The array was in the maxi-short 
configuration with the shortest baseline being 36~m. One primary flux and 
one polarized calibrator (as a pair) were observed at the beginning and 
end of each night.  For the primary flux calibrators we observed 3C147 and 
3C295, and for the polarized calibrators we observed DA240 and 3C303. We 
used the WSRT wide band correlator to cover a frequency range from 
310-390~MHz with eight 10~MHz wide bands, each with 64 channels and full 
Stokes parameters. We applied a Hanning taper to the spectral data, and 
every other channel was selected for analysis yielding an effective 
spectral resolution of about 0.31~MHz.  After removing the end channels in 
each band and editing for strong RFI, 197 channels remained in the final 
analysis, for a total bandwidth of 61~MHz.

The VLA observations were taken for 2.8 hours in December of 2005 in the D 
configuration, and the flux calibrators 3C48 and 3C286 were used. The data 
were taken using the spectrometer with no cross-had polarizations. We 
analyzed only the pseudo-continuum data for this analysis, with a 
frequency of 1.4649~GHz, a bandwidth of 12.5~MHz, and no polarization 
information.
 
\subsection{Total Intensity Calibration} The calibration and reduction of 
the WSRT and VLA data was performed using the NRAO's Astronomical Image 
Processing System (AIPS). For the WSRT data, the total intensity in each 
of the 8 bands was calibrated independently using standard procedures and 
the fluxes in the VLA calibrator manual for 3C147 and 3C295.  The VLA 
pseudo-continuum data set was calibrated using standard procedures. We did 
several iterations of amplitude and phase self-calibration on each data 
set.

\subsection{Polarization Calibration} Due to the fact that WSRT observes 
with orthogonal linear feeds (X and Y), the polarization calibration in 
classic AIPS involved several non-standard steps. WSRT polarization 
leakage terms are highly frequency dependent, so after the total intensity 
calibration, each channel was split into an independent uv-data set and 
the leakage terms were then solved for using the AIPS task LPCAL. Stokes Q 
values were calculated from (YY-XX)/2, while the Stokes U values are 
-(XY+YX)/2. An additional correction is needed for Stokes U, to remove the 
instrumental phase offset between the X and Y receivers using the 
polarized calibrators. Faraday rotation causes the Stokes Q and U 
amplitudes of the polarized calibrators to oscillate across the band. The 
X-Y phase offset was found for each channel by forcing the vector averaged 
Stokes $U=U\left(\lambda^{2}\right)$ visibilities for the polarized 
calibrators to match (offset by a quarter wavelength) the 
$Q\left(\lambda^{2}\right)$ sine wave formed by Faraday rotation. The 
observed sign of the Faraday rotation for 3C303 (+13~rad~m$^{-2}$, 
\citealt{kron77}) forced Stokes U values to be a quarter wavelength 
\emph{after} the Stokes Q values with increasing wavelength, breaking the 
final sign ambiguity and finishing the polarization calibration. However, 
we note that the intrinsic position angle of the polarization is still 
highly uncertain due to errors in the rotation measure, as discussed in 
$\S$3.3.

\section{Image Production \& Analysis} 

\subsection{Initial Spectral Cube} After calibration, we created I, Q, and 
U images for each of the 197 WSRT channels. All images and 
self-calibrations were done in AIPS with the tasks IMAGR and CALIB 
respectively. Each channel image was cleaned with IMAGR with 70,000 clean 
components and a gain of 0.1. The synthesized beam varied from 
$89\arcsec$$\times$$52\arcsec$ to $108\arcsec$$\times$$60\arcsec$ across 
the band, so all of the images were uv-tapered and restored to a common 
resolution of $108\arcsec$$\times$$60\arcsec$. Typical rms noise levels 
for the single-channel \{I,Q,U\} maps were \{1.0,0.5,0.5\}~mJy/beam, 
respectively.
     
\subsection{Total Intensity}

\subsubsection{WSRT I-Map} 
\begin{figure}[] 
\begin{center}
\includegraphics[width=12cm]{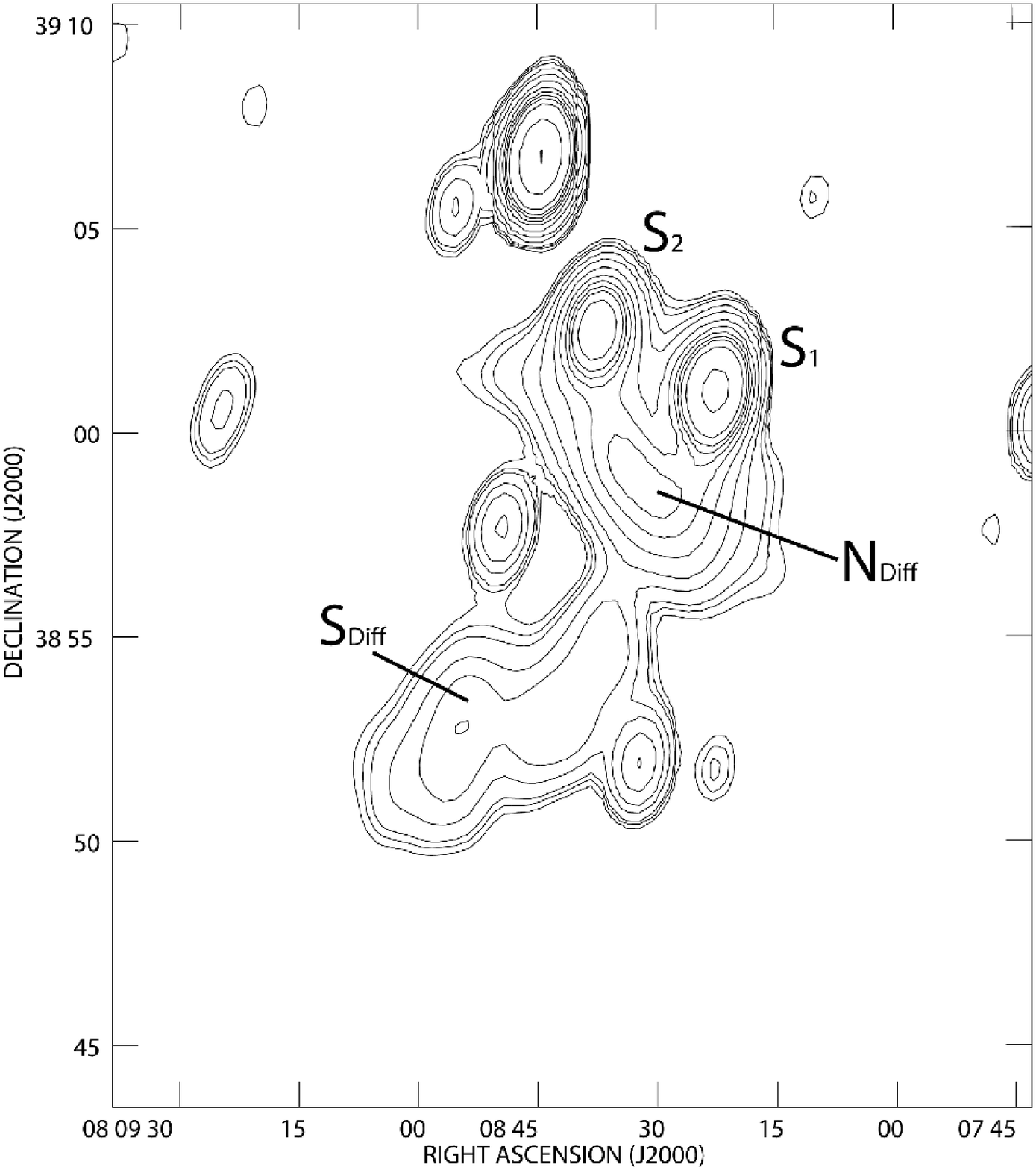}
\caption{\label{wsrti} WSRT 351 MHz total intensity image with 
contour levels 8.254$\times 10^{-3}$(-0.4, 0.4, 0.5, 0.8, 1.2, 2, 3, 4, 6, 8, 16, 32, 64) Jy/(108$\arcsec$$\times$60$\arcsec$ beam). S1 and S2 are ``compact" sources at 351 MHz that show sub-structure at higher resolution (see Figure \ref{vlai}).}
\end{center}
\end{figure}

The final WSRT I map is shown in Figure 
\ref{wsrti}. This map is the simple average of all the individual channel 
I maps ($108\arcsec$$\times$$60\arcsec$ beam). The average frequency is 
351 MHz and the observed noise is $\sigma\sim$188~$\mu$Jy/beam. The 
diffuse emission is seen to have two distinct components which we have 
labeled N$_{Diff}$ and S$_{Diff}$. At this resolution, N$_{Diff}$ 
partially blends into the compact sources $S_{1}$ and $S_{2}$.  As shown 
in $\S$3.2.2, some of these ``compact" sources also have substructure. 
Flux and size properties for both diffuse sources are summarized in Table 
1. We describe all the discrete radio sources in the next subsection.

\subsubsection{VLA I-Map} 
\begin{figure}[]
\begin{center}
\includegraphics[width=15cm]{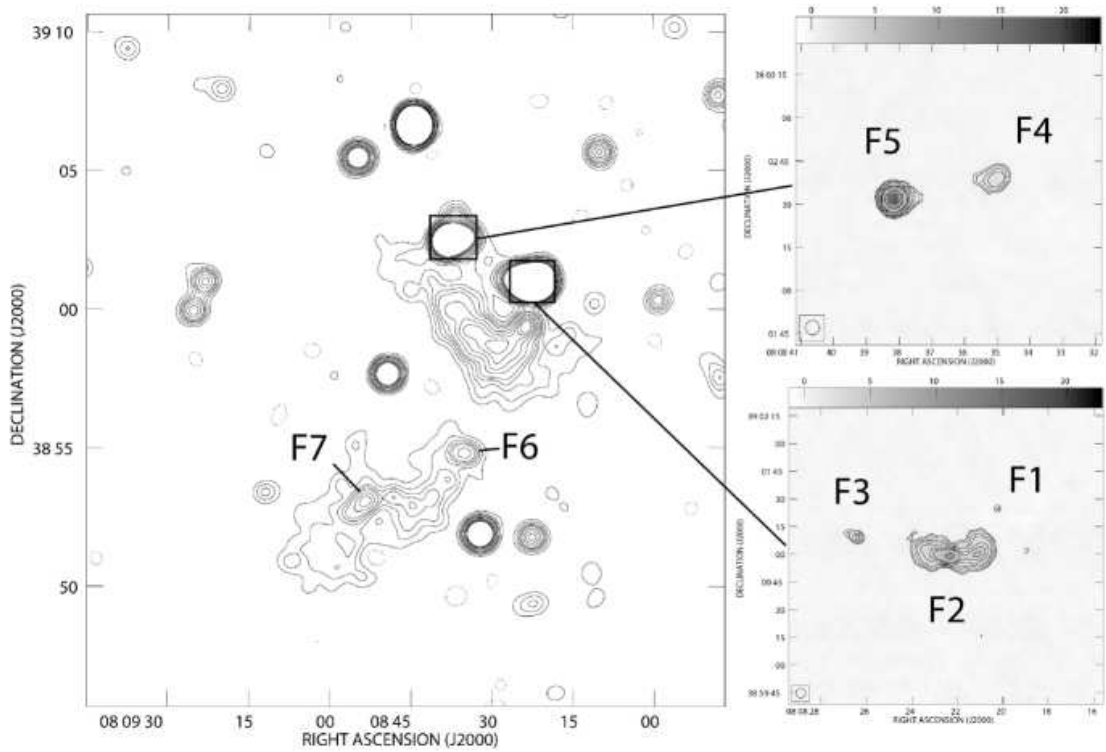}
\caption{\label{vlai} VLA I image (left) with contours 2.0$\times 10^{-4}$(-1, 1, 2, 
3, 4, 5, 7, 9, 11, 13, 15) Jy/(40$\arcsec$$\times$40$\arcsec$ beam). At right are close-ups views of S1 and S2 from the FIRST survey. Contours 4.0$\times 10^{-4}$(1.5, 2, 3, 4, 6, 8, 10, 20, 50, 100) Jy/(5$\arcsec$$\times$5$\arcsec$ beam).}
\end{center}
\end{figure}

Figure \ref{vlai} shows our VLA I map, which has 
a noise level of 90~$\mu$Jy/beam and a 40$\arcsec \times$40$\arcsec$ 
restoring beam. We have labeled the discrete radio sources, and have shown 
VLA FIRST \citep{beck95} images of some of the more interesting ones.  
The sources F1-F7 are adjacent to or embedded in the diffuse emission. F2 
is a 66 mJy wide-angle tailed (WAT: \citealt{owen76}) radio galaxy at 
z=0.196 (from SDSS:\footnote{http://www.sdss.org/} 
\citealt{york00,stou02}), and F3 and F5 are identified with SDSS galaxies 
with photometric redshifts of z=0.249 and z=0.255, respectively. In 
S$_{Diff}$, the source F6 is marginally detected in FIRST (not shown), and 
is coincident with a z=0.04 galaxy in SDSS. F7 is a FIRST radio point 
source (not shown) with no optical identification in SDSS.  The apparent 
bridge of emission between N$_{Diff}$ and $S_{1}$ peaks on a spiral galaxy 
at z=0.041 (see Figure \ref{sdssgray}) and is most likely diffuse disk 
emission from that galaxy.

\subsubsection{Spectral Index} 
\begin{figure}[]
\begin{center}
\includegraphics[width=10cm]{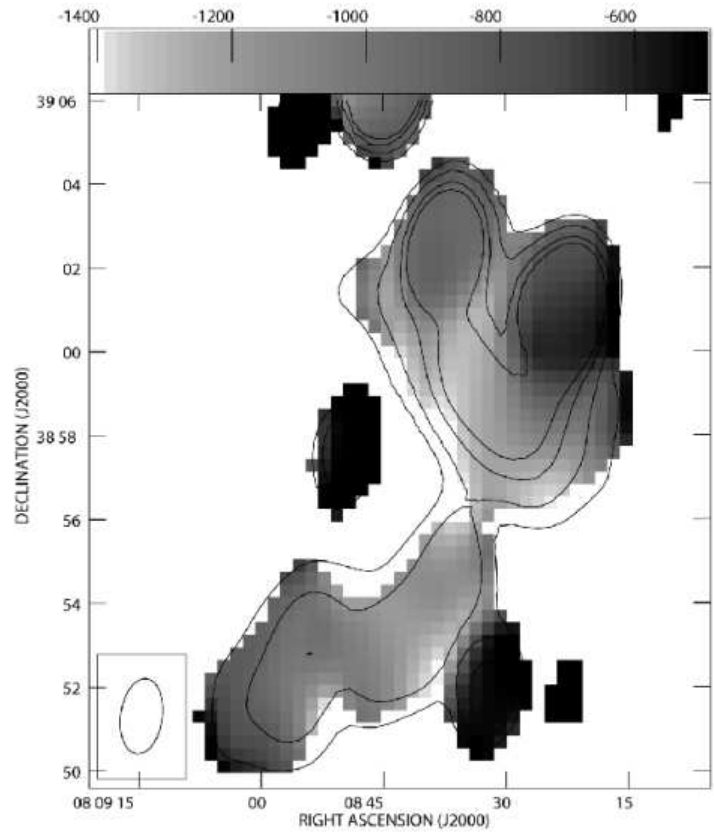}
\caption{\label{alpha} Spectral index map from 351 MHz to 1.4 GHz, with 
WSRT contours 8.25$\times 10^{-3}$(-0.5, 0.5, 0.8, 1, 2, 3, 6, 12, 
24, 48) Jy/(108$\arcsec$$\times$60$\arcsec$ beam).}
\end{center}
\end{figure}

For the spectral analysis we matched the 
uv-range of the VLA and WSRT data before comparison. The spectral index 
($\alpha$) map (Figure \ref{alpha}) was created in AIPS using the task 
COMB. Any pixel that was not at least 10$\sigma$ above the noise in either 
map was blanked. Both sources are steep spectrum, and Table 1 lists the 
integrated $\alpha$ obtained from fitting the total flux of each component 
at 1.4~GHz and 351~MHz to a power law.

\subsection{Polarization}

\subsubsection{Rotation Measure Synthesis} With the 197 channels and $\sim 
61$~MHz total bandwidth of our WSRT P-band data, it is possible to 
simultaneously determine the rotation measure distribution within each 
beam and remove the effects of bandwidth depolarization.  When searching 
for polarized synchrotron emission, which is often at very low surface 
brightness levels, one would like to observe over a large bandwidth 
($\Delta \nu$) to maximize the signal/noise ratio.  However, this normally 
results in depolarization from the vectoral cancellation of the Stokes Q 
and U signals that have been Faraday rotated from one side of the band to 
the other. Some radio telescopes allow for a large observing bandwidth to 
be split into many narrow channels (Westerbork being one of them), so an 
obvious solution would be to make an image of the polarized amplitude in 
each channel and average them together. Unfortunately, polarized galactic 
foreground emission is ubiquitous (e.g., \citealt{reic06}).  This is 
actually above the surface brightness of some diffuse extragalactic 
regions of interest, so adding the scalar polarization intensity of each 
channel will cause the galactic emission to add coherently as well. This 
increases the background and drowns out the desired diffuse extragalactic 
source.

\cite{bren05} presented a new technique to eliminate the confounding 
effects of intervening Faraday rotation (bandwidth depolarization and 
galactic foreground contamination) by searching ``rotation measure space" 
for polarized emission. The technique proceeds as follows: assume a 
rotation measure for the source, then derotate the polarization vector in 
each individual channel (to a reference wavelength $\lambda_{o}$) to 
correct for this and make a polarization map from the average of the 
derotated channels. If the source in fact had the assumed rotation 
measure, the channels would add coherently, allowing for the full 
sensitivity of the entire bandwidth.  The resulting map at a given Faraday 
depth, $\phi$, is approximately given by

\begin{equation}
F\left(\phi\right)=\frac{1}{N}\displaystyle\sum_{j=1}^{N}P_{j}e^{-2i\phi\left(\lambda^{2}_{j}-\lambda^{2}_{o}\right)}
\end{equation}

\noindent where N is the number of maps (channels) used and 
$P_{j}=Q_{j}+iU_{j}$ is the complex polarization at channel j. The units 
of $F\left(\phi\right)$ are Jy/beam/rmtf, where rmtf stand for the 
Rotation Measure Transfer Function. The RMTF is the response of a source 
with emission at a single Faraday depth. If one were able to sample an 
infinite range of $\lambda^{2}$, the RMTF would be a delta function in RM 
space, but incomplete sampling induces side-lobe structures in a manner 
similar to side-lobes in the beam pattern of an interferometer due to 
incomplete sampling of the uv-plane.

The key step is to apply Eq. (1) for a wide range of rotation measures, 
creating a rotation measure cube (RM-Cube), and search for coherent 
structures within this cube. We processed the 197 Q and U maps in IDL 
utilizing Eq. (1) to create the final RM-Cube, with Faraday depths running 
from -200~rad~m$^{-2}$ to +200~rad~m$^{-2}$ in steps of 1~rad~m$^{-2}$. 
This method has been successfully used to detect very diffuse polarized 
emission in the region of the Perseus cluster \citep{debr05}.

Near values of $\phi = 0$, polarized emission from our own galaxy fills 
the field of view. This emission has a typical surface brightness of 
$\sim$1~mJy/beam/rtmf, and in the vicinity of 0809+39 it peaks at 
$\phi_{Gal} \sim$+6~rad~m$^{-2}$.  The polarized emission in N$_{Diff}$ 
peaks near +12~rad~m$^{-2}$, though there is still a significant amount of 
galactic emission present at this Faraday depth. At values of $|\phi| > 
30~$rad~m$^{-2}$, very little of the galactic emission remains. As a 
result, the rms noise of the images decreases with increasing $|\phi|$, to 
$\sim$30~$\mu$Jy/beam/rmtf for $|\phi| >$100~rad~m$^{-2}$.

Some of the emission that \cite{debr05} detected in the field of the 
Perseus cluster with RM-Synthesis was later found to likely be Galactic in 
origin \citep{bren07}.  In the case of 0809+39, N$_{Diff}$, while not 
segregated significantly from the Galactic emission in Faraday depth, is 
an order of magnitude stronger in surface brightness, making a Galactic 
origin unlikely.

\begin{figure}[]
\begin{center}
\includegraphics[width=9cm]{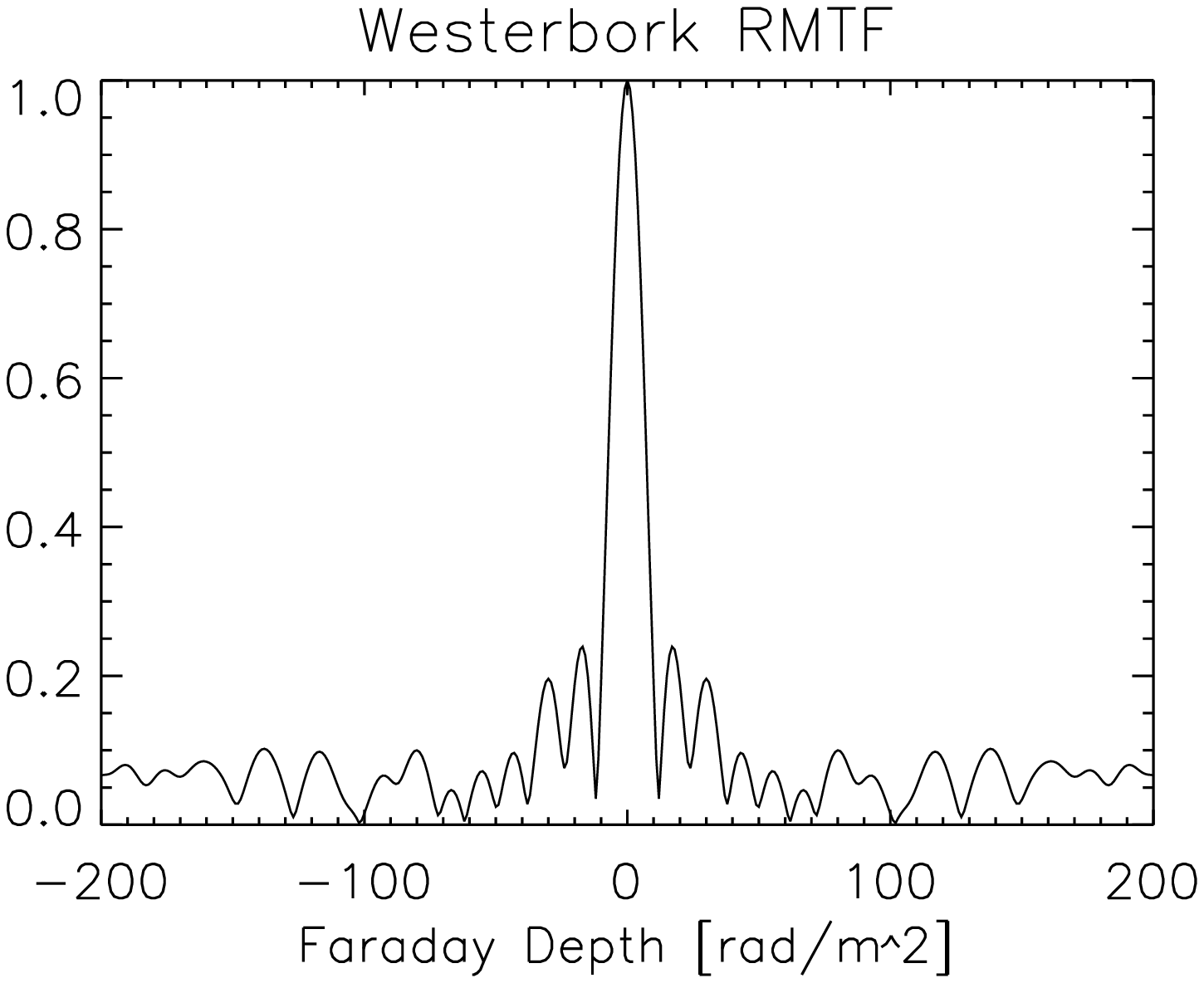}
\includegraphics[width=9cm]{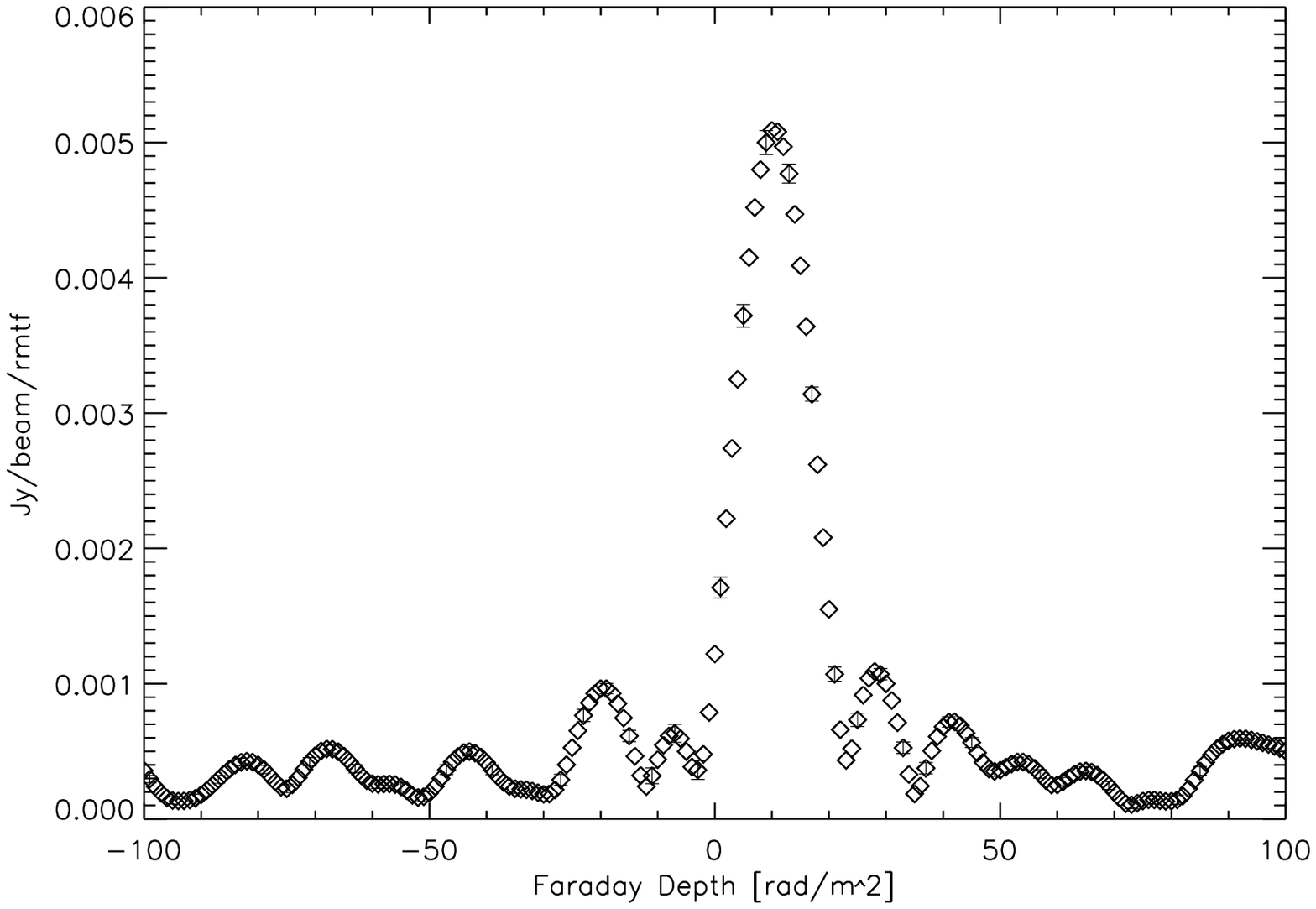}
\includegraphics[width=9cm]{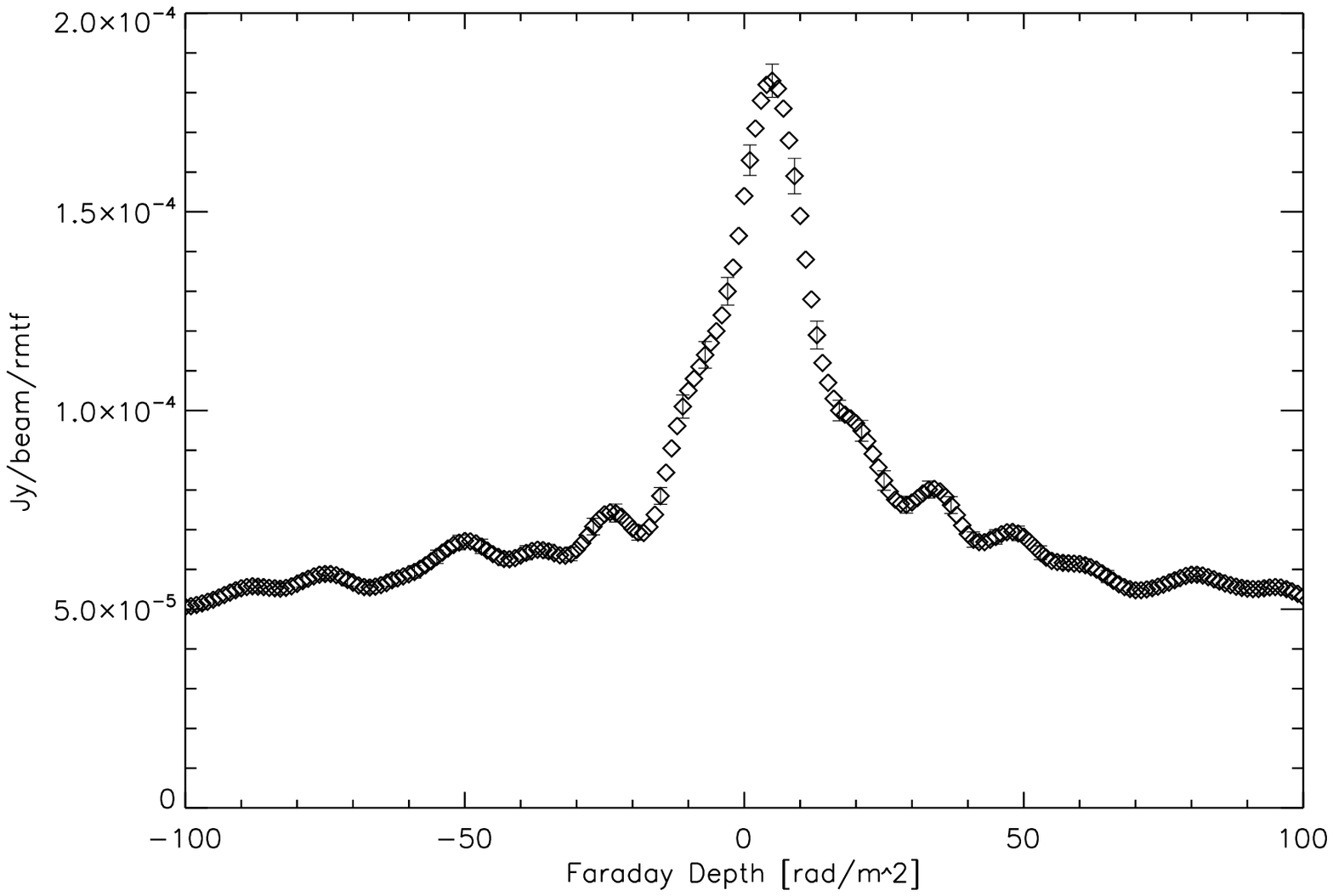}
\caption{\label{fspec} Top: Our RMTF; Center: Faraday spectrum for N$_{Diff}$; Bottom: Faraday spectrum for a large area of the Galactic emission. Characteristic error bars are shown.}
\end{center}
\end{figure}

We can also use this RM-Cube to find Faraday spectra, which is just 
$F\left(\phi\right)$ from Equation 1 for a single pixel or region in the 
sky. Figure \ref{fspec} shows the average spectrum of a 
$1.8\arcmin$$\times$$1.8\arcmin$ region centered on N$_{Diff}$, along with 
the RMTF for our frequency sampling. The Faraday spectrum of N$_{Diff}$ is 
very close to a ``point source" in Faraday space, especially when compared 
to a typical Galactic spectrum also shown in Figure \ref{fspec}. The 
diffuse galactic radiation along the line of sight to 0809+39 has emission 
at multiple Faraday depths. This galactic signal must also be present in 
our spectrum of N$_{Diff}$, but is lower than the side-lobe level of 
N$_{Diff}$ itself.

\subsubsection{Position Angle} 
\begin{figure}[]
\begin{center}
\includegraphics[width=10cm]{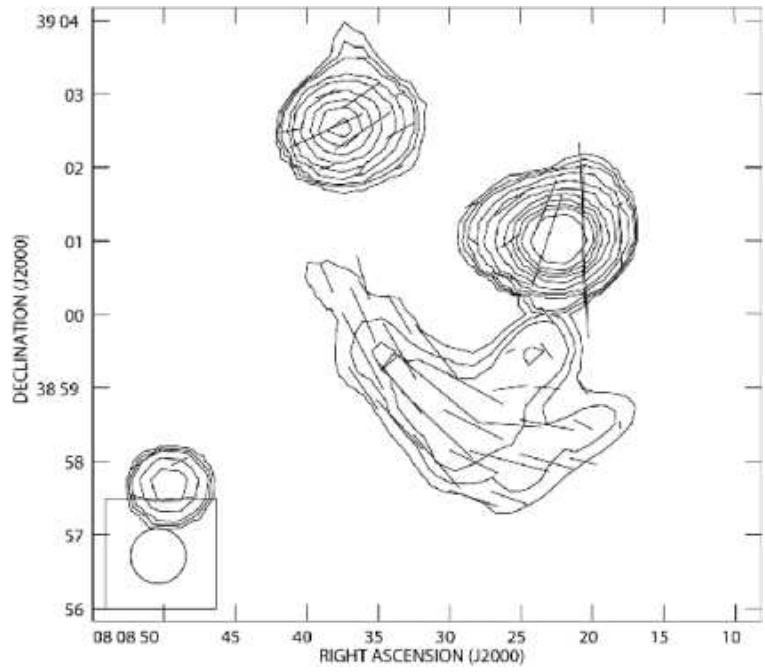}
\caption{\label{nvsspol} B field orientation obtained from derotating 
the NVSS to correct for a rotation measure of +12~rad~m$^{-2}$. Contours are NVSS I 
at 4.9$\times 10^{-3}$(-0.3, 0.3, 0.4, 0.5, 0.8, 1.2, 2, 3, 4, 5, 6, 8, 
16, 32, 64)~Jy/(45$\arcsec$$\times$45$\arcsec$ beam), 
and the magnetic field lines are 1$\arcsec$=2.22$\times 10^{-5}$~Jy/(45$\arcsec$$\times$45$\arcsec$ beam).} \end{center}
\end{figure}

We used the NRAO VLA Sky Survey (NVSS: 
\citealt{cond98}) polarization image of 0809+39, along with our measured 
RM, to find the absolute position angle $\chi$ of N$_{Diff}$. Figure 
\ref{nvsspol} shows NVSS total intensity contours and magnetic field 
orientation (which is $\chi \pm 90^{o}$), corrected for the average $\phi$ 
of N$_{Diff}$. Due to the difference in resolution between the RM-Cube and 
the NVSS, and the fact that the gradient in $\phi$ across the source 
translates into only a $\delta \chi \sim 7^{o}$, we did not do a 
pixel-by-pixel de-rotation. The magnetic field runs along the long axis of 
N$_{Diff}$, and shows evidence of following the curvature in the southern 
end.

\begin{figure}[]
\begin{center}
\includegraphics[width=10cm]{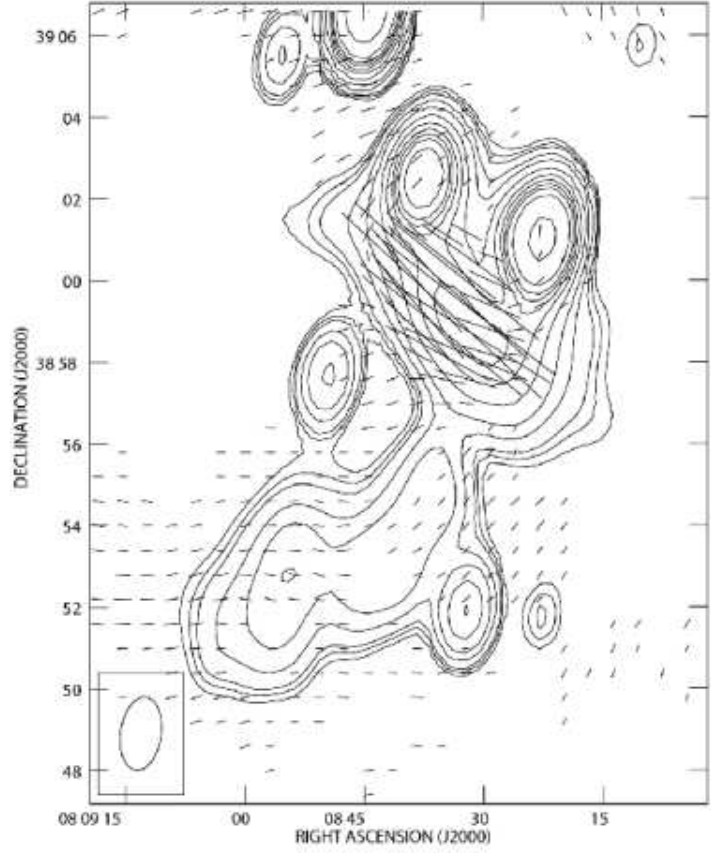}

\caption{\label{wsrtpol} WSRT 351 MHz contour image of 0809+39 I (same 
contours and resolution as Figure \ref{wsrti}) with corrected magnetic 
field vectors for the RM-Cube at $\phi = +12~rad~m^{-2}$. 1$\arcsec$ = 1.85$\times 
10^{-5}$~Jy/(108$\arcsec$$\times$60$\arcsec$ beam). The apparent ``noise" 
in the polarized emission is actually real emission from our own galaxy.}

\end{center}
\end{figure}

Our Faraday corrected WSRT magnetic field vectors are plotted in Figure 
\ref{wsrtpol}. We obtained the polarization from the complex RM-Cube at 
$\phi=+12$~rad~m$^{-2}$, where N$_{Diff}$ peaks. Given the uncertainties 
in absolute position angle when derotating from $\lambda$=90~cm to 
$\lambda$=0 (e.g., $\sigma_{\chi} \approx 140^{o}$ when 
$\sigma_{\phi}=3$~rad~m$^{-2}$), we globally rotated the vectors so that 
the magnetic field orientation of N$_{Diff}$ matches that seen in Figure 
\ref{nvsspol}. We placed the polarized intensity cut-off such that some of 
the diffuse Galactic emission can be seen. One can see the discontinuity 
in angle between the galactic polarization and the brighter N$_{Diff}$ 
emission.

\subsubsection{Fractional Polarization} The detection of N$_{Diff}$ in the 
NVSS survey allows for examination of the degree of de-polarization.  The 
fractional polarization, m, decreases from $m \sim 45-55$\% at 21cm 
(matched to the WSRT resolution) to $m \sim 20$\% at 92~cm. Assuming for a 
moment that this depolarization is due only to internal Faraday 
de-polarization \citep{burn66,ciof80}, we calculate the necessary internal 
Faraday rotation to be $\phi_{in} \approx 2$~rad~m$^{-2}$. The upper limit 
intrinsic width of N$_{Diff}$ is $\approx 7-9$~rad~m$^{-2}$, so the 
depolarization could be due to internal Faraday de-polarization. At 
351~MHz, S$_{Diff}$ is not polarized at a 3$\sigma$ level of 9\%.

\section{X-ray \& Optical Identification} We now turn our attention to the 
optical and X-ray environment of 0809+39. Since all radio halo and relic 
sources found thus far exist in or adjacent to the hot gas of galaxy 
clusters, we searched for thermal emission from X-ray clusters in the 
vicinity of 0809+39 and investigated the surrounding optical field.

\subsection{X-Ray} 
\begin{figure}[]
\begin{center}
\includegraphics[width=15cm]{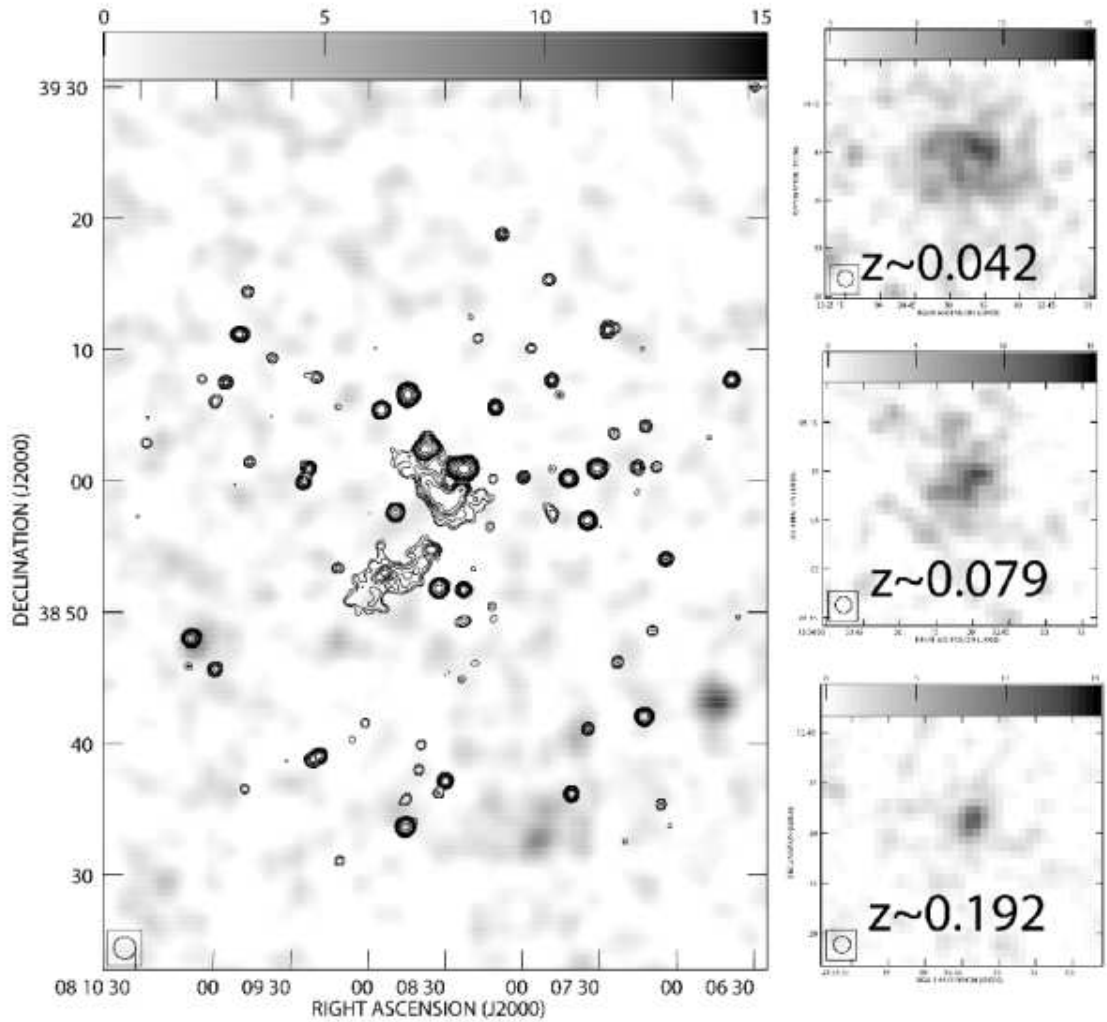}
\caption{\label{rosat} ROSAT broad (0.1-2.4 keV) continuum grayscale 
(convolved to 100$\arcsec$) with VLA L-band contours. To the right are ROSAT 
images of three X-ray selected clusters at the indicated redshifts, at the 
same grayscale and resolution as the 0809+39 ROSAT image.}  
\end{center}   
\end{figure}

Figure \ref{rosat} shows the ROSAT broad (0.1-2.4 keV) 
continuum emission in the region of 0809+39 with VLA L-band radio contours 
overlayed. To put the low observed brightness of the X-ray field of 
0809+39 in context, we have also plotted three X-ray selected clusters at 
three different redshifts. The clusters are \{RXCJ2324.3+1439, 
RXCJ1353.0+0509, RXCJ2155.6+1231\} with redshifts of \{0.042, 0.079, 0.192 
\} and X-ray luminosities of L$_{X}$=\{0.97, 1.97, 5.51\}$\times 
10^{44}~h^{-2}_{70}$~erg~s$^{-1}$, respectively \citep{pope04}. The 
cluster redshifts were selected to mimic the relevant systems we have 
identified in $\S$4.2. It is clear that no diffuse X-ray emission like 
that present in the X-ray selected clusters is present in the ROSAT 
0809+39 field.

\begin{figure}[]
\begin{center}
\includegraphics[width=7cm]{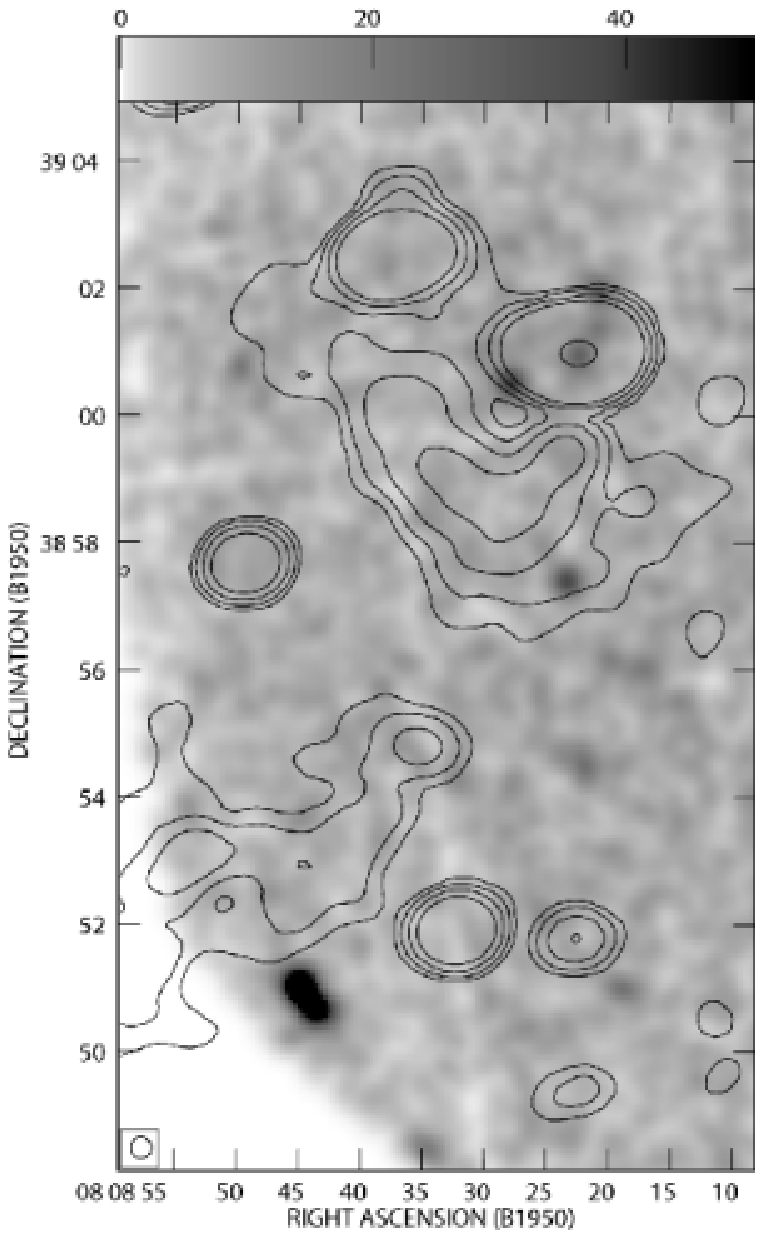}
\includegraphics[width=9cm]{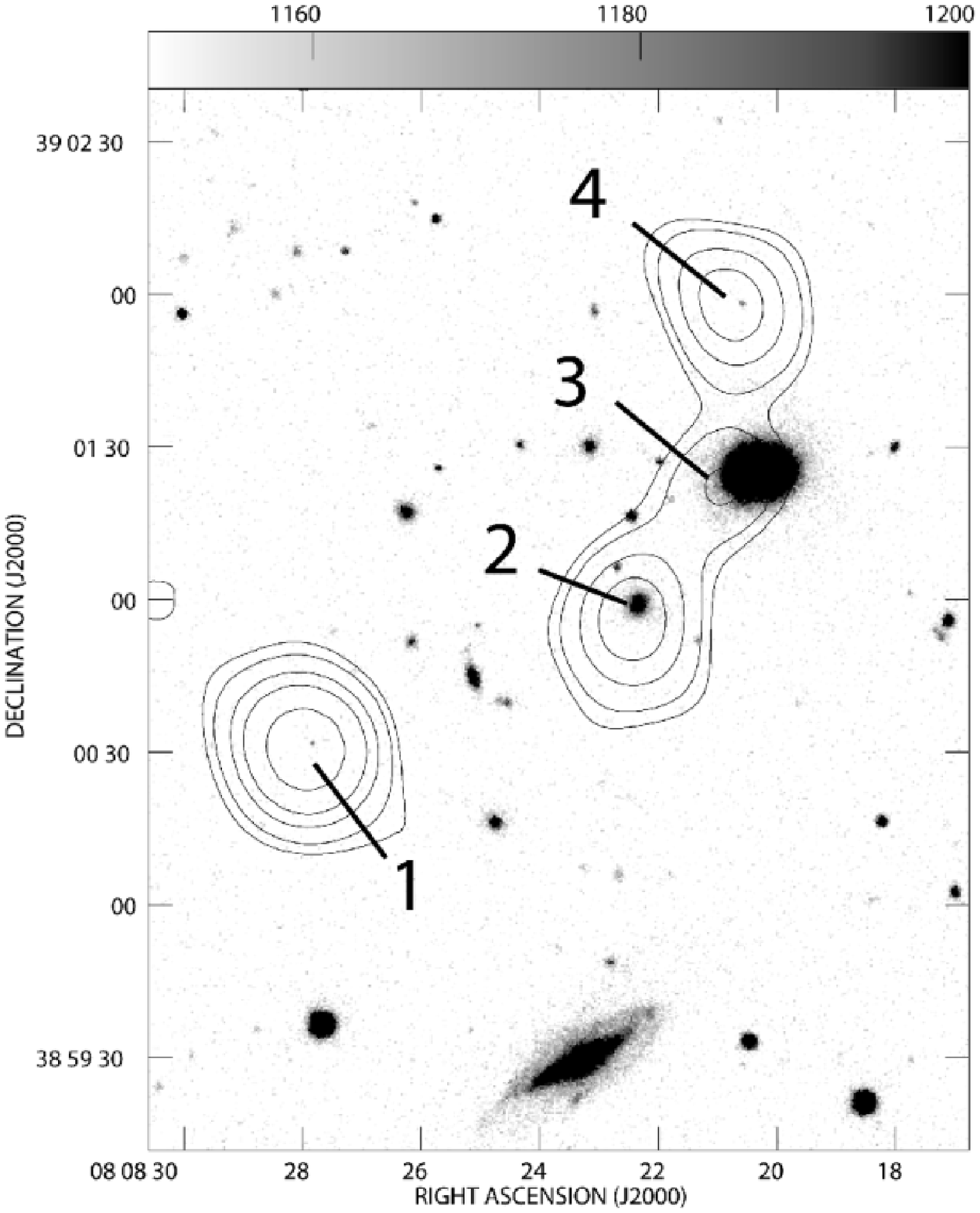}
\caption{\label{xmmsmall} Left:  XMM Epic observation in counts (convolved 
to 20$\arcsec$) with VLA radio contours 6.5$\times 10^{-3}$(0.05, 0.1, 
0.2, 0.4, 8)~Jy/(49$\arcsec$$\times$42$\arcsec$ beam). Note that a faint 
X-ray source is coincident with F2 (the WAT seen in Figure \ref{vlai}); 
Right: SDSS R image with XMM EPIC contours. Grayscale is in units of 
counts, and the contours are 4.08$\times$(3.5, 4, 5, 6, 8) counts. X-ray 
source 2 is the WAT (F2) and the galaxy near source 3 is coincident with 
compact radio source F1.}
\end{center}
\end{figure}

\begin{figure}[]
\begin{center}
\includegraphics[width=15cm]{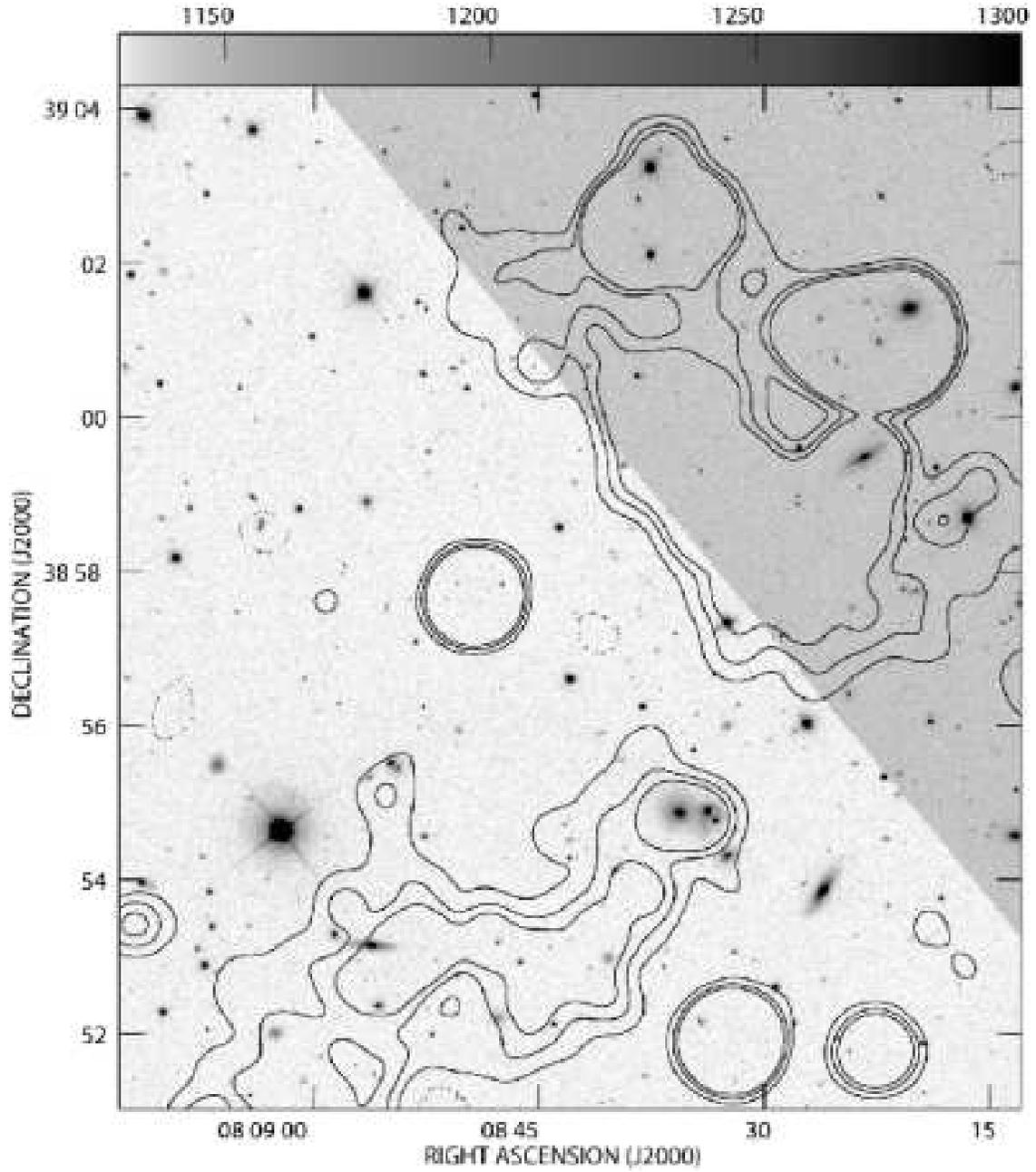}
\caption{\label{sdssgray} SDSS R mosaic grayscale (in counts) with VLA 
1.4 GHz contours at 2.0$\times$10$^{-4}$~(-1, 1, 2, 
3)~Jy/(40$\arcsec$$\times$40$\arcsec$ beam).}
\end{center}
\end{figure}

Figure \ref{xmmsmall} shows VLA contours over XMM EPIC grayscale as well 
as XMM contours over SDSS R grayscale from an XMM observation of the 
nearby galaxy UGC 4229 (P.I. Matteo Guainazzi) where 0809+39 was toward 
the edge of the field. There is clear emission from the WAT, but there are 
also several peaks in the region of N$_{Diff}$. Peak 2 is located on the 
WAT. Peak 1 is coincident with an SDSS photometric object at $z=0.313\pm 
0.1$, and peak 4 is coincident with an SDSS photometric object at 
$z=0.272\pm 0.06$. Peak 3 is offset $\sim 10''$ from a 2MASS galaxy at 
$z=0.02$, which is also the FIRST source F1 in Figure \ref{vlai}.

After subtracting out the four point sources, we found no evidence of 
excess diffuse emission. Using 
webPIMMS\footnote{http://heasarc.nasa.gov/Tools/w3pimms.html} we calculate 
a 3$\sigma$ upper limit X-ray luminosity of L$_{X}$(0.1-2.4~keV)=1$\times 
10^{43}$~erg~s$^{-1}$ at an assumed redshift of z=0.2.

\subsection{Optical} As one would expect from a region that spans more 
than $400\arcsec$ on the sky, the optical field of 0809+39 contains 
multiple, overlapping redshift systems. This is illustrated by Figure 
\ref{sdssgray}, which shows an SDSS mosaic R image with VLA contours. To 
set the scale, the large spiral galaxies in this image are at a redshift 
of roughly z~=~0.04. We begin by outlining the known and cataloged optical 
systems in this region, then focus our attention on the systems we believe 
are most likely associated with N$_{Diff}$ and S$_{Diff}$.

\begin{figure}[]
\begin{center}
\includegraphics[width=10cm]{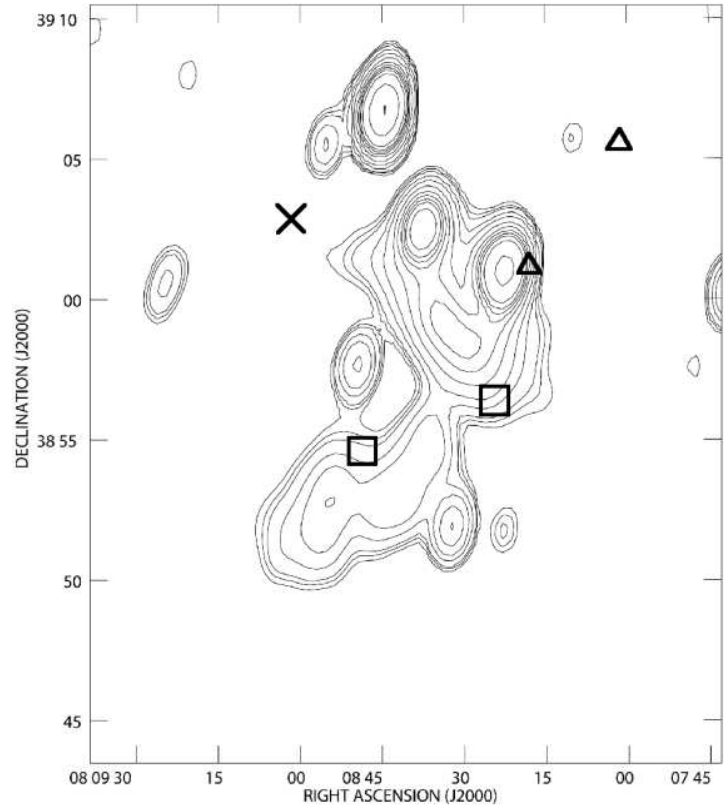}
\caption{\label{knownclust} Plot of known groups/clusters in the 0809+39 
region. Boxes are z$\approx$0.04 systems, triangles are z$\approx$0.07 systems, 
and the cross is a z$\approx$0.11 cluster.}
\end{center}
\end{figure}

\cite{dela06} suggested an association of both N$_{Diff}$ and S$_{Diff}$ 
with a grouping of galaxies at z$\sim$0.04. A $z=0.04063$ group (16 
members) is reported by \cite{mill05} from the SDSS cluster catalog C4, 
and \cite{merc05} find a group at z=0.040346 (19 members) from DR3. 
\cite{gal03} also detected z$\sim$0.073 and z$\sim$0.11 clusters in this 
region. These are plotted in Figure \ref{knownclust}. From Figures 
\ref{rosat} and \ref{xmmsmall} one can see that none of these have 
associated X-ray emission. 

\begin{figure}[] 
\begin{center}
\includegraphics[width=11cm]{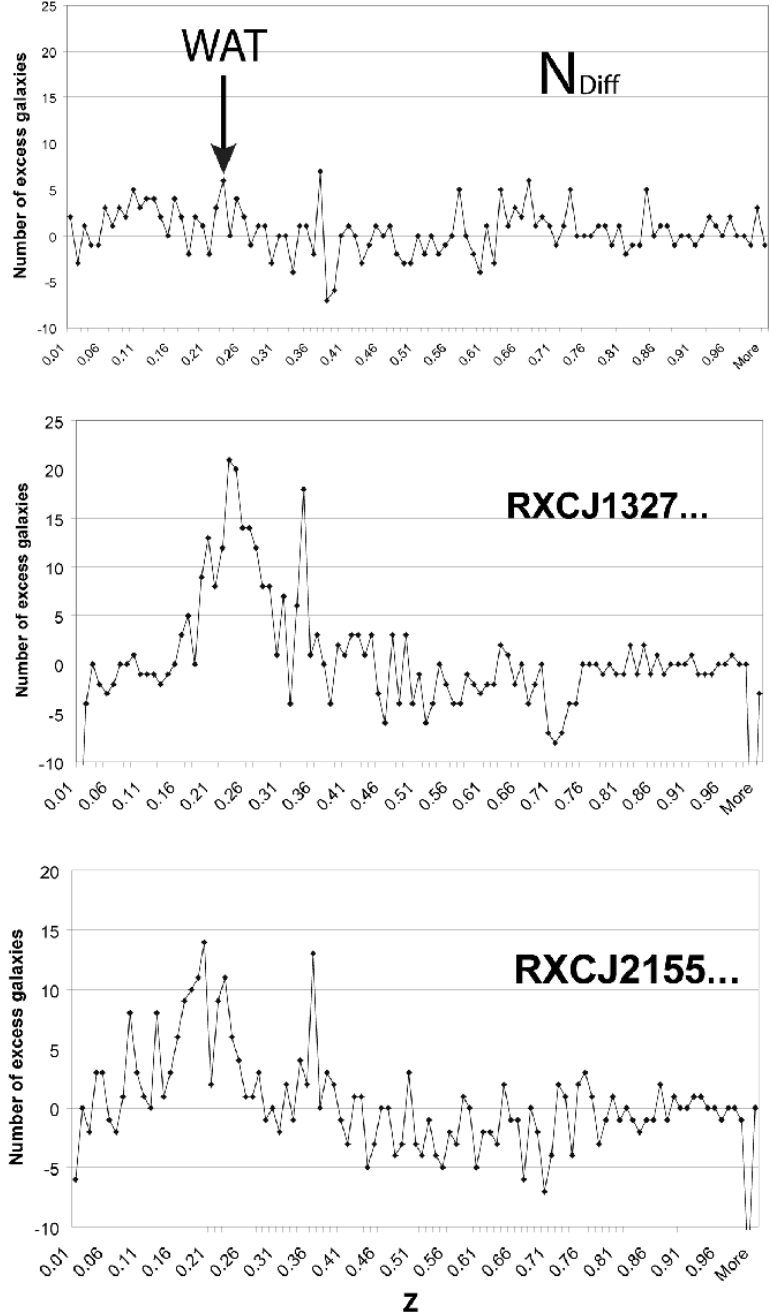}
\caption{\label{histo0809} Histograms of galaxy photometric redshifts
from SDSS for 6$^{\prime}$ around the WAT source F2 (Top, see Figure 
\ref{vlai}) and the X-ray selected clusters RXCJ1327.0+0211 (Center) and 
RXCJ2155.6+1231 (Bottom) (Popesso et al. 2004). Note that all the histograms have a narrow peak at z$\sim$0.35. This is an artifact of the template fitting method for calculating the photometric redshifts (\citealt{csab03}; see Figure 16 in that paper).}
\end{center}
\vspace{-.4in}
\end{figure}

\begin{figure}[] 
\begin{center}
\includegraphics[width=5cm]{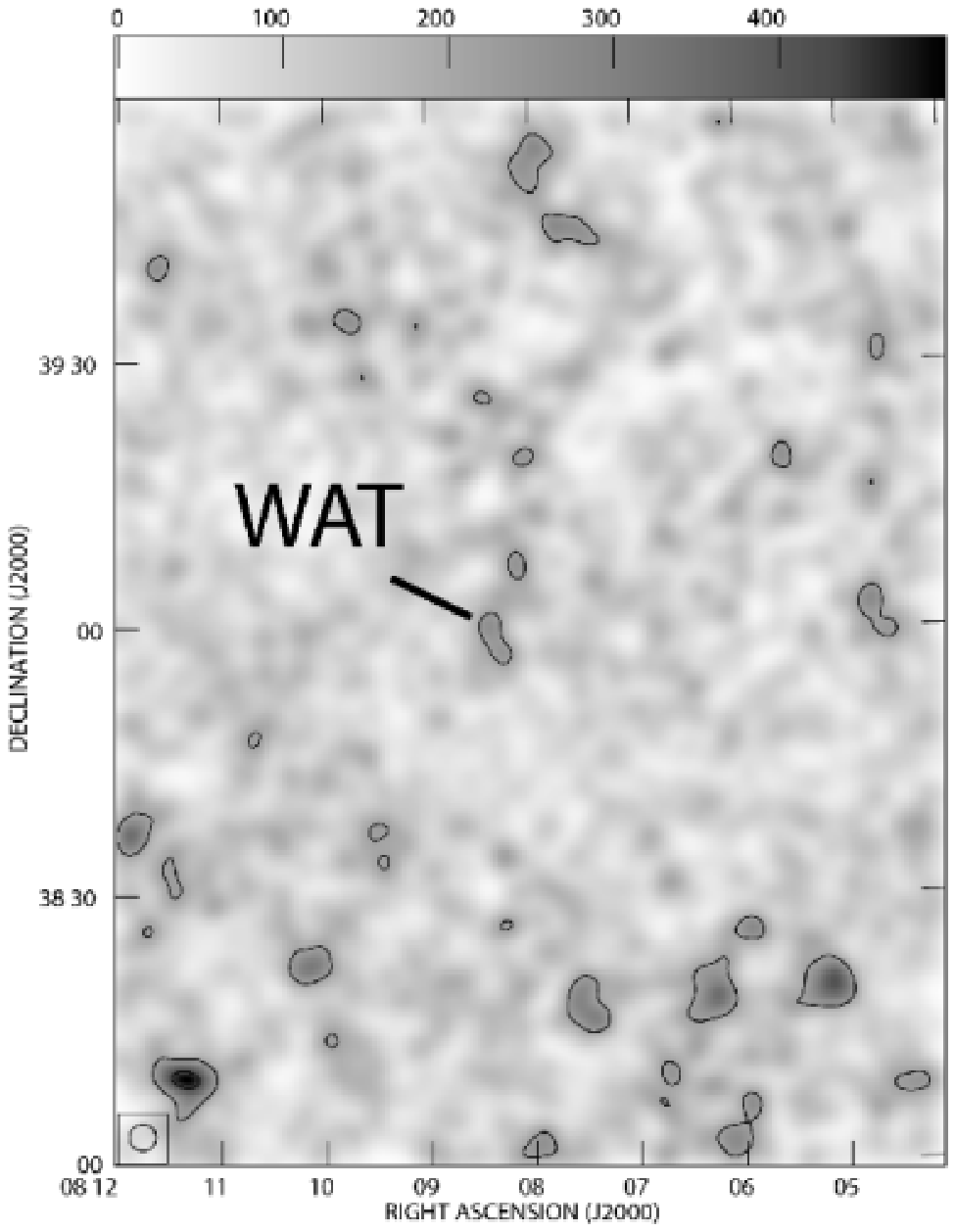}
\includegraphics[width=5cm]{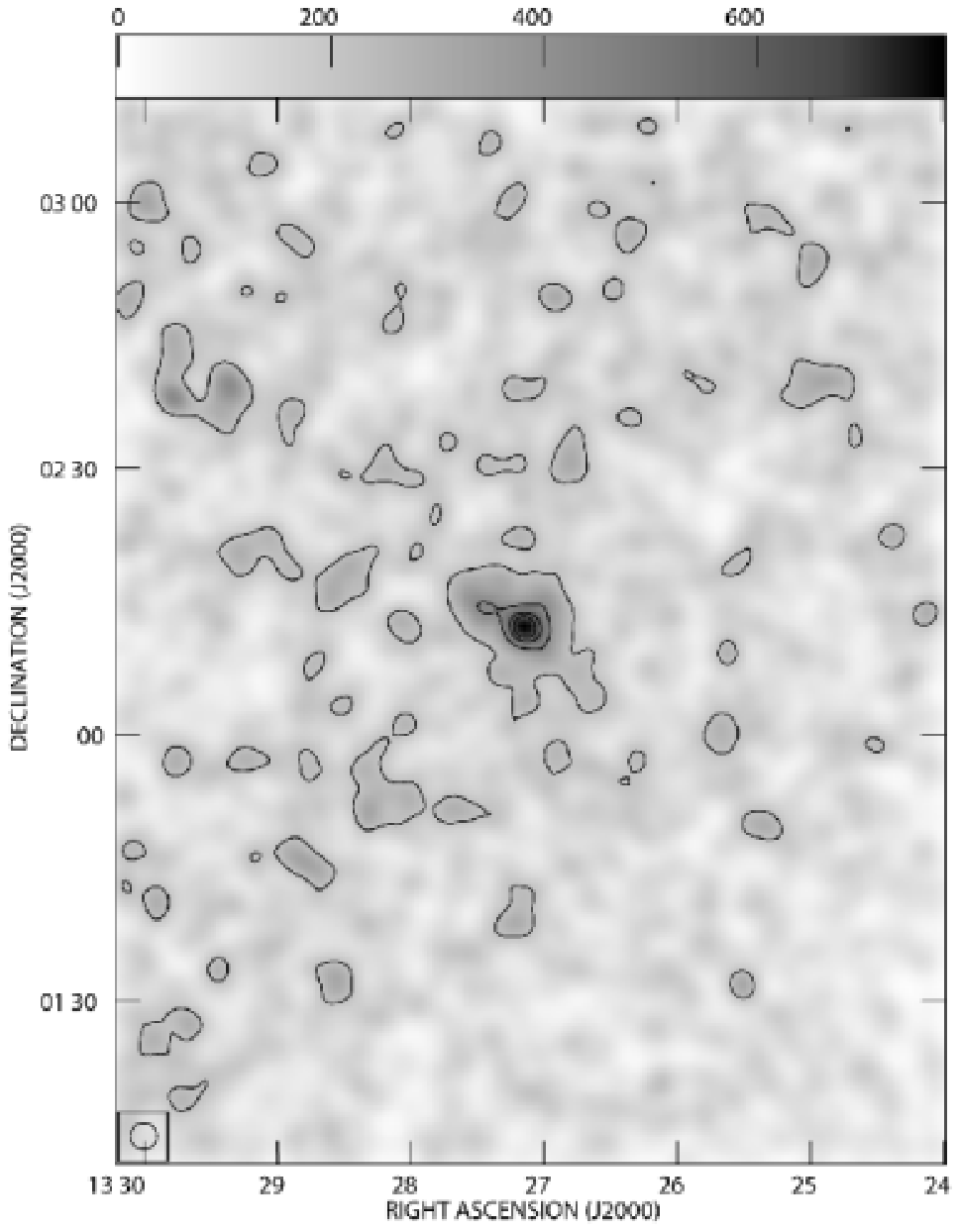}
\includegraphics[width=5cm]{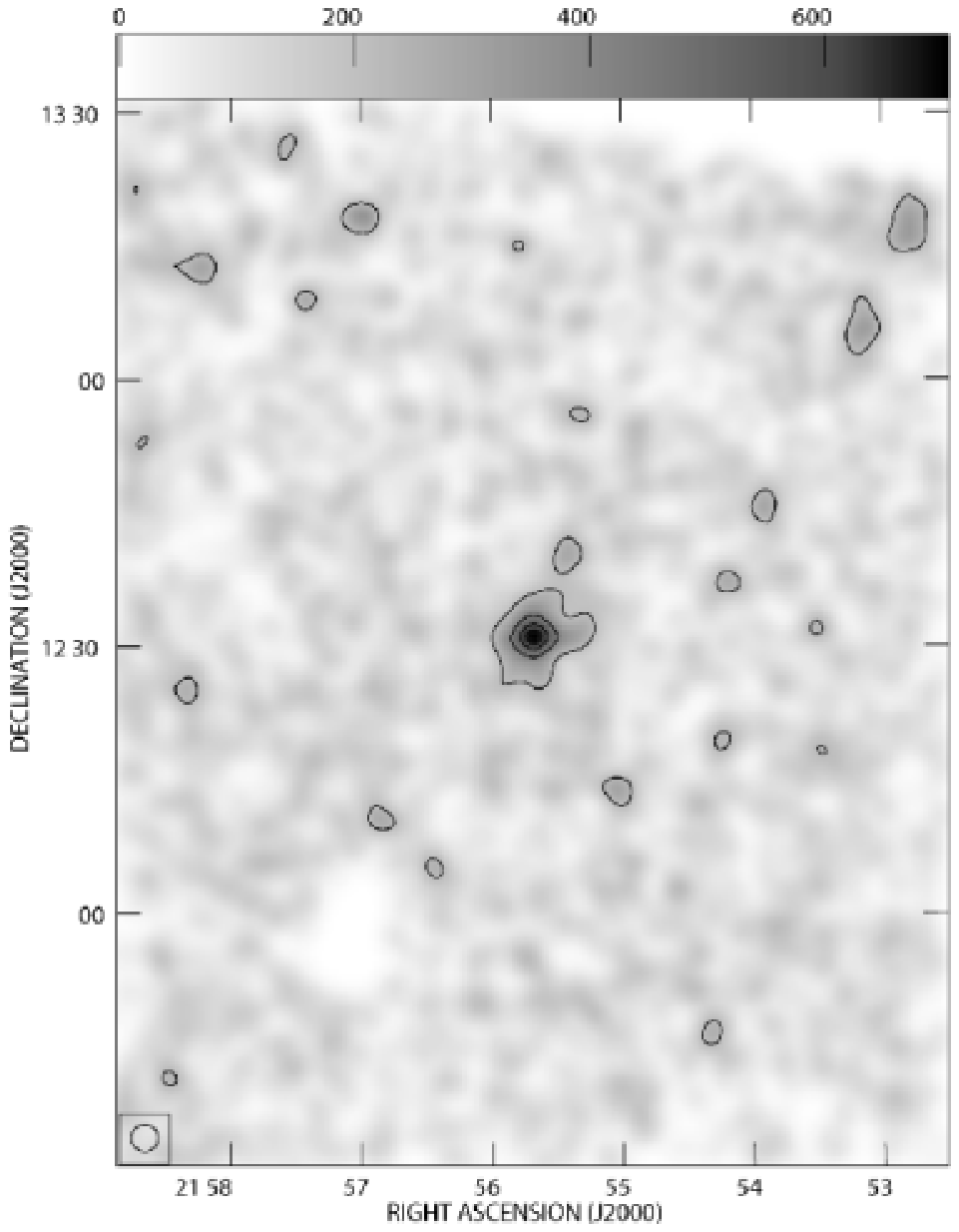}
\caption{\label{sdss0809} Distribution of galaxies from the SDSS photometric 
database with redshifts between 0.15 $<$ z $<$ 0.25, smoothed to 6$^{\prime}$. 
Left: 0809+39; Using the same redshift limits we show two X-ray selected clusters for comparison: Middle: RXCJ1327.0+0211; Right: RXCJ2155.6+1231}
\end{center}
\vspace{-.4in}
\end{figure}

Due to the presence of the WAT at z=0.2, we consider this redshift system 
as likely associated with N$_{Diff}$ (see arguments in $\S$5.1). To 
assess the significance of the clustering at this redshift, we used the 
SDSS photometric galaxy database \citep{adel07} and made a histogram in 
redshift of all the galaxies within a radius of 6$^{\prime}$ ($\sim$1~Mpc 
at z=0.2) from the WAT. We then subtracted a histogram (normalized to the 
same number of total counts) of a roughly 2$\times$2 degree field around 
the WAT, in order to subtract out any systematics in SDSS's photometric 
redshifts. The results are displayed in Figure \ref{histo0809}. Figure 
\ref{sdss0809} shows the spatial distribution of galaxies (number of 
galaxies/pixel at the redshift of the WAT z=0.2$\pm$0.05) from the SDSS 
photometric data. A weak clustering of galaxies is seen around the WAT. We 
also show in Figures \ref{histo0809} and \ref{sdss0809}, for comparison, 
the results from two X-ray selected clusters at similar redshifts 
(discussed in $\S$5.1). To assess the relative richness of this 
group/cluster, we followed the analysis used in \cite{gal03} for 
estimating the richness of clusters found in the POSS-II survey. We took 
the number of galaxies (above a background value determined from the 
surrounding 2x2 degree field) within $\sim$1~Mpc with an absolute R 
magnitude of -22.53 $<$ M$_{R}$ $<$ -19.53 taken from the SDSS photometric 
data. The grouping of galaxies associated with the WAT has a richness of 
16, which is on the poorer end of the overall distribution of richnesses 
in DPOSS \citep{gal03}.  We should note that the cluster at z=0.073 
detected by \cite{gal03} is also very close to N$_{Diff}$, and is just as 
likely to be associated with the diffuse emission as the WAT grouping of 
galaxies, but the issues discussed in $\S$5 apply to either case. The 
cluster at z=0.073 had a richness of 24.9, also significantly poor, and 
roughly corresponding to an Abell richness class of R$<$0 
\citep{abel89,gal03}.

\begin{figure}[] 
\begin{center}
\includegraphics[width=10cm]{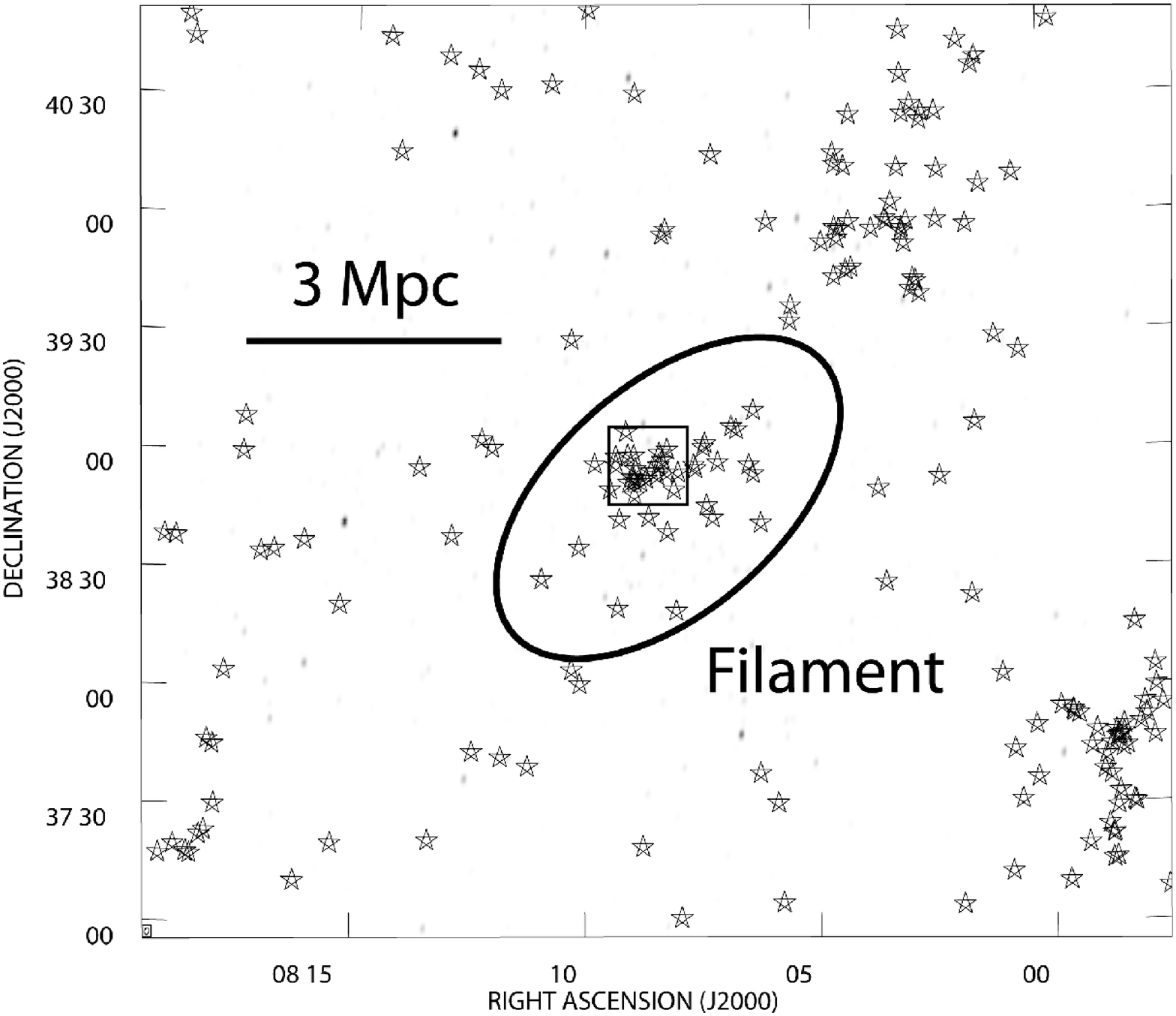}
\includegraphics[width=10cm]{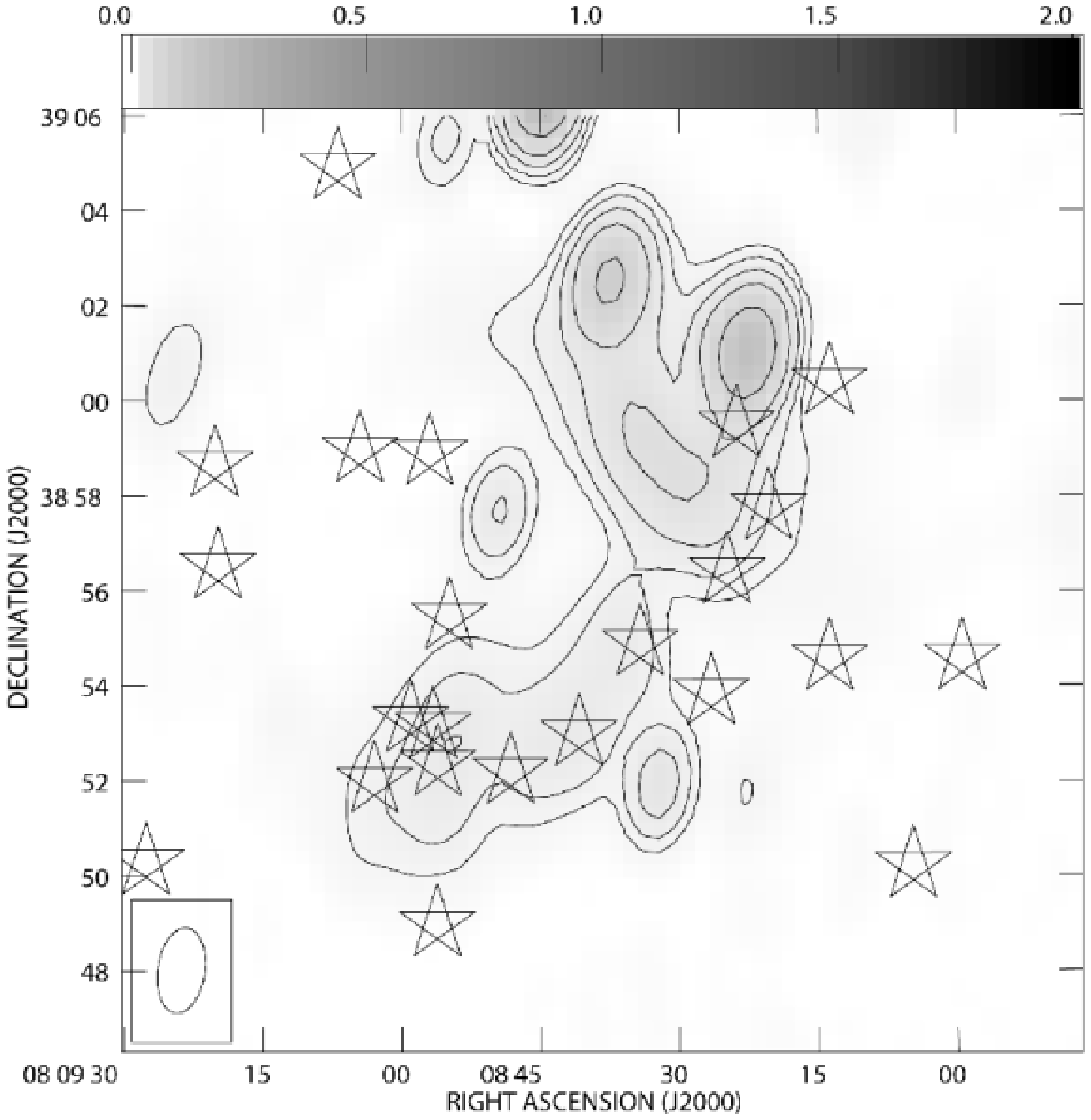}
\caption{\label{spec_z_0809} Top: Distribution of 0.0366 $<$ z $<$ 0.0448 
galaxies with spectroscopic red-shifts in SDSS plotted as stars. Bottom: 
close-up view of the boxed region in the top panel. S$_{Diff}$ is 
embedded in the filament of galaxies.} \end{center} 
\vspace{-.4in}
\end{figure}

\begin{figure}[] 
\begin{center}
\includegraphics[width=10cm]{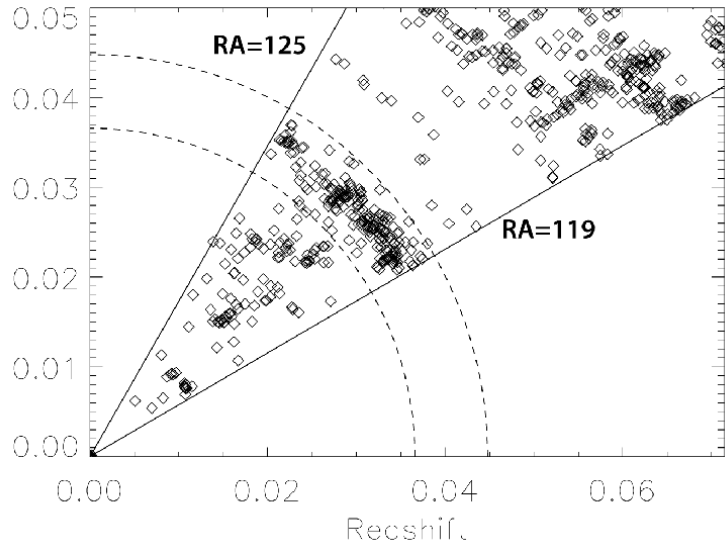}
\caption{\label{spec_dist} Distribution of redshift vs. RA for galaxies 
with spectroscopic redshifts in SDSS in a roughly 4$\times$4 degree field 
around 0809+39. The dashed lines enclose the range in z used to create 
Figure \ref{spec_z_0809} (0.0366 $<$ z $<$ 0.0448), and show the 
clustering of the filament galaxies. There are other significant 
structures at higher redshifts, but the filament at z=0.04 was deemed 
significant based on its spatial correlation with S$_{Diff}$.}
\end{center} 
\end{figure}

If we allow for the proximity of S$_{Diff}$ with N$_{Diff}$ to be 
coincidental, then identifying the optical system that is physically 
connected with S$_{Diff}$ is not straightforward. The three likely 
possibilities are the z$\sim$0.04, 0.073, and 0.2 groupings of galaxies. 
We visually searched the spatial distribution of galaxies from the SDSS 
photometric catalog to find a redshift where there was significant 
clumping in or near S$_{Diff}$, without success. The spectroscopic 
database of galaxies in SDSS, while much sparser, does reveal a 
$\sim$5~Mpc filament of galaxies surrounding S$_{Diff}$ at redshifts of 
0.0352 $<$ z $<$ 0.0441 (Figure \ref{spec_z_0809}). Many of these galaxies 
are those that make up the two groups previously detected in this region 
\citep{mill05,merc05}, but looking at a wider field reveals the larger 
filament. We performed the same richness analysis on the filament 
(centered on RA=122.224, DEC=38.883, 1~Mpc radius) as we did with the WAT 
group, and found a richness of 13. Figure \ref{spec_dist} shows a wedge 
diagram of redshift vs. RA for SDSS galaxies with spectra in a roughly 
4$\times$4 degree field around 0809+39. We have indicated the range of 
galaxy redshifts that are plotted in Figure \ref{spec_z_0809}. Groupings 
of galaxies at higher redshift were not correlated spatially with 
S$_{Diff}$.

\section{Physical Origin of the Radio Emission} We now turn our attention 
to the physical origin of the diffuse emission.  Though we cannot strictly 
rule out a Galactic origin for either source, the emission is not 
morphologically similar and is far brighter than typical diffuse galactic 
emission in this region. We hereafter assume that both components are of 
extragalactic origin. There are many different types of extragalactic 
large-scale diffuse radio emission \citep[e.g.,][]{kemp04}, but they in 
general fall within one of two basic classes: 1) Those directly powered by 
current AGN activity; 2) Those associated with processes in the 
intracluster medium (ICM). We include in the second class emission related 
to processes in the intergalactic plasma of galaxy filaments 
\citep{kim89,giov90,bagc02,kron07}.  In the vast majority of cases, 
determining the identification is straightforward. Either the diffuse 
region is directly connected (via jets or filamentary bridges) to an 
active radio galaxy, or there is a rich cluster of galaxies nearby whose 
potential well (highlighted by $\sim$10$^{8}$~K, X-ray emitting gas) 
provides an obvious energy source for particle acceleration (via 
gravitational collapse/accretion --$>$ shocks/turbulence etc).

Both sources in 0809+39 are unique in that neither of these conditions apply 
(see \citealt{dela06} for two similar sources). There are no cataloged X-ray 
emitting clusters nearby, only groups and poor clusters of galaxies. An AGN 
origin for these sources is also not obvious. In this paper we narrow our 
analysis to the two most likely (or least improbable) sources for the radio 
emission.

\subsection{Northern Component} In this section, we: 1) Rule out 
N$_{Diff}$ being an extended radio lobe directly powered by the WAT; 2) 
Give evidence that N$_{Diff}$ is a classical ``radio relic"; 3) Compare 
N$_{Diff}$ with other relics and conclude that it has an anomalously high 
radio/X-ray luminosity ratio; 4) Determine that the X-ray luminosity is 
appropriate for this optical richness, and is thus not to blame for the 
abnormal luminosity ratio; 5) Examine possible sources of the relativistic 
electrons such as direct acceleration from the thermal plasma, adiabatic 
compression (only) of fossil WAT plasma, and reacceleration of fossil WAT 
plasma. We conclude that the most likely source is reacceleration of 
fossil plasma from past WAT activity.

We first consider whether N$_{Diff}$ could be lobe emission from an AGN. 
The nearby WAT (F2) is the only reasonable candidate as an AGN source. 
However, they are spatially well separated, with a peak to peak distance 
$>$500~kpc at z=0.2. While the FWHM of the long axis of N$_{Diff}$ is 
200$\arcsec$, there are faint wings that extend out to 600$\arcsec$. This 
corresponds to $\sim$2 Mpc at z=0.2, which is a size comparable to that of 
a giant radio galaxy (GRG). However, its morphology is unlike any other 
GRG known \citep[e.g.,][]{mach01}. We can also consider the lifetime of 
particles emitting at 1.4~GHz. The minimum energy magnetic field for 
N$_{Diff}$ is B$_{min} \sim$ 0.6~$\mu$G, which fixes the Lorentz factor at 
$\gamma \sim 2\times10^{4}$ and results in inverse Compton losses 
dominating the lifetime.  From \cite{sara99}

\begin{equation}
t_{IC}=2.3\times 10^{12}\gamma^{-1}\left(1+z\right)^{-4}~yr,
\end{equation}

\noindent which for N$_{Diff}$ is $t_{IC}\sim 10^{7.7}$~yr. Relaxing the 
minimum energy requirement for magnetic field strength, we can calculate 
the maximum lifetime for any electron radiating at 1.4~GHz following the 
prescription of \cite{sara99}. Assuming a redshift of z=0.2 yields a 
maximum lifetime of t$_{max} \sim 10^{8}$~yr, similar to that calculated 
above using B$_{min}$.

If we assume a 1~keV gas temperature and only hydrogen gas the sound crossing 
time is t$_{sc} \sim 10^{9.1}$~yr. Therefore the timescale for 1.4~GHz IC 
losses will be much shorter than the diffusion/sound-crossing time (from the 
WAT to N$_{Diff}$). If the WAT was the original source for the N$_{Diff}$ 
plasma, e.g. from an earlier outburst, then there must have been some 
re-acceleration or adiabatic enhancement. This is no surprise because this has 
been a longstanding result for radio halos and relics.

There are several lines of evidence that point toward N$_{Diff}$ being a 
radio ``relic" source associated with a poor cluster at z$\sim$0.2: 1) 
Both the clumping in redshift and the existence of the WAT \citep{owen76} 
indicate the presence of a cluster; 2) WATs are also known to be 
associated with merger dynamics 
\citep{pink93,gome97,roet96,loke95,pink00,blan03}. Additional supporting 
evidence for merger activity near N$_{Diff}$ comes from the multiple X-ray 
peaks seen in Figure \ref{xmmsmall}, one of which is coincident with a 
z=0.27 SDSS galaxy. Currently all known radio relic or halo sources have 
been found in/near clusters in a disturbed dynamical state \citep{fere06}; 
3) The long axis of the diffuse emission is perpendicular to the line 
connecting N$_{Diff}$ and the WAT, typical of relic sources; 4) The 
diffuse emission is highly polarized, also typical of radio relic sources 
(e.g. \citealt{giov04}); 5) The magnetic fields are parallel to the long 
axis of the emission, suggesting shock compression.

 All of these point toward N$_{Diff}$ being either a Radio Phoenix or 
Radio Gischt \citep{kemp04}, depending on whether the ``seed" plasma came 
from an extinct radio galaxy lobe or was initially accelerated at a 
cluster accretion shock, respectively. The curvature of the WAT (from 
Figure \ref{vlai}) is toward the North, suggesting infall from the South. 
With a longest linear extent of $\sim$2 Mpc, N$_{Diff}$ is comparable to 
larger relic sources around rich clusters of galaxies \citep{giov99}. For 
radio relics where the spectral index distribution is known, the edge 
farthest from the cluster is always sharper and has a flatter spectrum 
\citep{giov04}. Though N$_{Diff}$ does not exhibit this behavior (see 
Figure \ref{alpha}), we should note that the source is only a little more 
than a single WSRT beam thick in the transverse direction. Therefore a 
higher resolution spectral index map is needed to confirm this.

\begin{figure}[]
\begin{center}
\includegraphics[width=10cm]{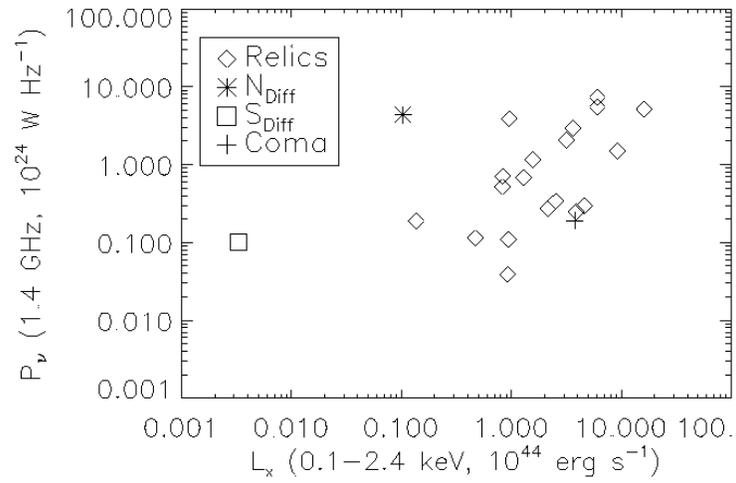}
\caption{\label{lrvslx} Plot of relic radio luminosity/Hz at 1.4 GHz vs. the 0.1-2.4 keV
X-ray luminosity of the associated cluster. Compiled are 22 radio
relics from Giovannini et al. (1991); Giovannini, Tordi, \& Feretti
(1999); Kempner \& Sarazin (2001); Govoni et al. (2001); Slee et al.
(2001); Govoni et al. (2005). They represent a complete list of radio
relics that have reasonably reliable 1.4~GHz Flux density measurements.}
\end{center} 
\vspace{-.4in}
\end{figure}

We can now ask how N$_{Diff}$ compares to other observed radio relics. 
Radio halos, diffuse radio emission centered on some clusters of galaxies, 
are known to exhibit a good correlation between their radio luminosity and 
the X-ray luminosity of their associated cluster 
\citep[e.g.,][]{fere03,fere06}. A similar but weaker correlation for radio 
relics has been claimed by \cite{giov04}, who quoted $P_{1.4GHz} \propto 
10^{K} L_{x}$, where K ranged from 0.8 to 2.2. We have compiled a 
relatively complete list of known relics with available 1.4~GHz flux 
measurements and plotted their radio luminosities vs. X-ray luminosities 
in Figure \ref{lrvslx} \citep[compiled 
from][]{giov91,giov99,kemp01,govo01,slee01,govo05}. N$_{Diff}$ is at least 
an order of magnitude too luminous (under-luminous) in the radio (X-ray). 
We examine several possible explanations as to why the radio or X-ray 
emission from the relic source N$_{Diff}$ is not what we would expect from 
the radio/X-ray luminosity relation for rich galaxy clusters.  From the 
observed correlation, N$_{Diff}$ should have an $L_{x} \sim 
10^{44-45}~erg~s^{-1}$ from 0.1-2.4~keV, but from XMM observations we 
measure a 3$\sigma$ upper limit of $L_{x} \approx 1\times 
10^{43}$~erg~s$^{-1}$. We examine the two quantities in this relation 
separately, starting with the X-ray luminosity.

It is possible that the grouping of galaxies that N$_{Diff}$ is associated 
with is massive enough to emit the expected amount of X-rays (i.e. $\sim 
10^{44-45}$~erg~s$^{-1}$ at 0.1-2.4~keV), but for some reason they are not 
observed. Perhaps the grouping will have the ``correct" X-ray luminosity 
eventually, but we have caught it in a very early evolutionary state where 
the thermal gas has not reached the needed density or temperature to emit 
sufficiently in the X-rays. The optical properties of the WAT group 
($\S$4.2), however, are not representative of massive X-ray emitting 
clusters (see, e.g., \citealt{ledl03} for optical vs. X-ray properties of 
Abell clusters). To further show this, we examined the optical properties 
of the X-ray selected\footnote{We used X-ray selected clusters to avoid 
pre-selecting optically rich clusters.} clusters RXCJ1327.0+0211 and 
RXCJ2155.6+1231, both of which have a similar redshift \{0.259, 0.192\} to 
the WAT and a L$_{x}$ =\{1.67, 1.12\}$\times10^{45}$~erg~s$^{-1}$, 
respectively \citep{pope04}. For the optical data we again used the SDSS 
photometric galaxy database and performed the same analysis as with 
the WAT system ($\S$4.2). Figure \ref{histo0809} shows the histograms and 
Figure \ref{sdss0809} the distribution of galaxies with the same contour 
levels as the WAT system. The X-ray clusters show much stronger clustering 
than the WAT system, both spatially and in redshift. Using the same 
richness analysis that we used for the WAT group, RXCJ2155.6+1231 and 
RXCJ1327.0+0211 have richnesses of 42 and 53, respectively, consistent 
with large X-ray luminous clusters (e.g. \citealt{ledl03}). The lack of 
X-ray emission in the WAT group, therefore, is consistent with its poor 
optical properties, and cannot explain the discrepant radio/X-ray 
luminosity ratio seen in Figure \ref{lrvslx}.
  
We now examine the radio luminosity, which is apparently two orders of 
magnitude too luminous given the observed correlation for radio relics. We 
proceed under the assumption that N$_{Diff}$ is related to the presence of 
a shock, as we argued earlier, and attempt to understand the source of the 
relativistic electrons that are causing the synchrotron emission. The two 
likely possibilities are that the electrons were accelerated directly out 
of the thermal plasma by diffusive shock acceleration (DSA), or the seed 
electrons came from fossil plasma from past AGN activity and were 
re-accelerated and/or adiabatically enhanced by the shock.

\cite{mini01} performed a cosmological simulation that included only DSA 
of cosmic rays from the thermal environment and did not include fossil AGN 
plasma. They report a correlation between 1.4~GHz radio luminosity from 
primary electrons (those accelerated at shocks) and cluster temperature 
(see Figure 6 of that paper) that is roughly consistent with the observed 
correlation for radio relics shown in Figure \ref{lrvslx}. The radio/X-ray 
luminosity ratio of N$_{Diff}$ is therefore inconsistent with the results 
of \cite{mini01}. We can examine the conditions under which their 
simulations would have produce the observed P$_{\nu}$/L$_{X}$ ratio of 
N$_{Diff}$. Using observed L$_{X}$ vs. T$_{cluster}$ relations 
\citep{hart08} and theoretical expectations for low-density environments 
\citep{ryu03}, we assume a T$_{cluster}$ $<$ 1 keV. \cite{mini01} would 
then predict P$_{\nu}$(1.4~GHz) $\approx 1\times 10^{22}$~W~Hz$^{-1}$, 
assuming an acceleration efficiency of 10$^{-4}$ and an electron to proton 
injection ratio of R$_{e/p}$=0.01. Therefore, either the acceleration 
efficiency needs to be $\sim 0.01$ or R$_{e/p} \sim$1 in order for 
\cite{mini01} to reproduce the P$_{\nu}$/L$_{X}$ ratio of N$_{Diff}$. It 
is thus physically possible to use DSA out of the thermal gas to create 
N$_{Diff}$, but it is not clear why the efficiency or R$_{e/p}$ should be 
so anomalous in this region.

The other possible source of seed electrons is fossil or ``relic" plasma 
from past AGN activity, presumably from the nearby WAT. There are two 
possible mechanisms for reviving old plasma. The shock can either 
adiabatically enhance the relativistic particles and magnetic fields of 
the relic plasma \emph{only} \citep[e.g.,][]{enss01}, or the shock can 
reaccelerate the particles as well.

In the case of pure adiabatic compression, let us compare the energy 
content of the current WAT radio lobes and N$_{Diff}$. From the FIRST 
data, the eastern lobe of the WAT has a minimum energy magnetic field of 
B$_{min} \approx$ 2.7~$\mu$G and a total energy of E $\approx 6 \times 
10^{57}$~erg. N$_{Diff}$ has B$_{min} \approx$ 0.6~$\mu$G and E $\approx 
3.4 \times 10^{59}$~erg. It appears that the energy contained in 
N$_{Diff}$ cannot be explained by adiabatically compressing an extinct 
radio lobe similar to the current lobes of the WAT.  This does not rule 
out this scenario however, since the current activity of the WAT may not 
be indicative of past activity, and WATs in general are known to have 
energies of the same order as N$_{Diff}$ \citep[e.g.,][]{deyo84}.

The reacceleration of the fossil electrons can further increase the 
emissivity of the relic plasma \citep{blan87,mico99,mark05}, potentially 
by an order of magnitude, depending on the pre-shock spectral index and 
the shock compression ratio R. From the spectral index of N$_{Diff}$, 
$\alpha = -1.12$, we can find the shock compression ratio R = 
$\frac{\alpha-1}{\alpha + 1/2}$ = 3.4 \citep{bell78,drur83} and Mach 
number M $\approx$ 2.1. The Mach number is a reasonable one for this 
environment \citep[e.g.,][]{ryu03}. In short, energetically it is not out 
of the question that the radio emission N$_{Diff}$ was created by shock 
compressed fossil plasma from the WAT, assuming that the WAT's past 
activity was stronger than it is currently.

Given that DSA of electrons from the thermal plasma reproduces the 
observed radio vs. X-ray luminosity correlation for radio relics, from 
which N$_{Diff}$ is a clear outlier, we conclude that reacceleration 
and/or adiabatic compression of fossil plasma from the WAT source is a 
more likely origin for the radio emission in N$_{Diff}$. Due to the fact 
that current activity of the WAT is not energetically comparable to 
N$_{Diff}$, we must invoke more powerful activity in the past in order to 
make the adiabatic compression/reacceleration hypothesis work. Had we 
associated N$_{Diff}$ with the z$\sim$0.07 group of galaxies ($\S$4.2, 
Figure \ref{knownclust}), the P$_{\nu}$/L$_{X}$ ratio would still have 
been two orders of magnitude too large for the observed correlation for 
radio relics. Our arguments would have proceeded the same way, except we 
would have needed to invoke past activity from an \emph{undetected} AGN to 
explain the discrepant radio luminosity.

\subsection{Southern Component}

\begin{figure}[]
\begin{center}
\includegraphics[width=10cm]{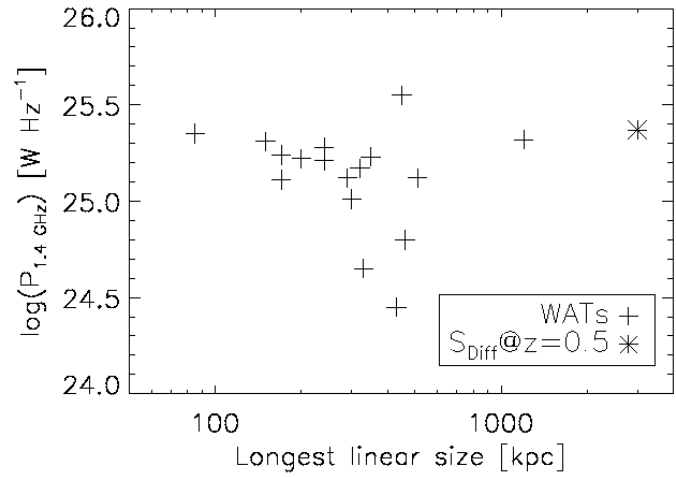}
\caption{\label{watpowersize} 1.4 GHz power vs. linear extent in kpc for
a sample of WAT sources (Pinkney et al. 2000). S$_{Diff}$, if at redshift
of z$>$0.5, is an extremely large radio galaxy.}
\end{center}
\end{figure}

Many of the same issues that we found with N$_{Diff}$ arise when we 
consider possible origins for S$_{Diff}$. If we consider an AGN origin, 
the morphology of S$_{Diff}$ (Figure \ref{vlai}) is reminiscent of a WAT 
source centered on the FIRST radio source F7. F7, however, does not have 
an optical counterpart in SDSS. If it is an unidentified WAT, its redshift 
is likely to be z $>$ 0.5 \citep{schm06}, making its total linear extent 
$>$3~Mpc. Figure \ref{watpowersize} shows a plot of 1.4~GHz radio power 
vs. linear extent for a sample of WAT sources from \cite{pink00}. If 
S$_{Diff}$ were indeed at a redshift of z $>$ 0.5, it would have a linear 
size that far exceeds typical WATs in this sample. It is also possible 
that one of the z=0.04 galaxies that makes up the filament (Figure 
\ref{spec_z_0809}) hosts an AGN that created the extended emission, but 
none of the galaxies at this redshift (with spectra in SDSS) show signs of 
AGN activity. However, the AGN could have been disrupted and disappeared 
leaving the lobe emission behind \citep[e.g.,][]{parm07}, as long as this 
happens on a timescale less than the 1.4 GHz electron maximum lifetime of 
10$^{8}$~yr (see discussion in $\S$5.1). Assuming S$_{Diff}$ is at a 
redshift of z=0.04, its total energy would be E~$\approx 7.9\times 
10^{57}$ erg and the minimum energy magnetic field would be B$_{min} 
\approx 0.6~\mu$G, not atypical for WAT lobe emission (see $\S$5.1).

S$_{Diff}$ could be caused by ICM or IGM processes, similar to our claim 
for N$_{Diff}$.  From Figure \ref{rosat} we can see that, like N$_{Diff}$, 
there is no cluster X-ray emission detected in either ROSAT or XMM.  The 
apparent filament in which S$_{Diff}$ is embedded offers an intriguing 
origin for the diffuse emission. Unlike other diffuse radio sources that 
have claimed to be part of filamentary large-scale structure 
\citep{kim89,giov90,kron07}, S$_{Diff}$ is not near \emph{any} massive 
clusters. At 390~kpc long, S$_{Diff}$ is smaller than the $\sim$1.5~Mpc 
emission in the Coma-Abell 1367 supercluster \citep{kim89,giov90} or the 
$4-5$~Mpc radio regions found by \cite{bagc02} and \cite{kron07}. 
\cite{bagc02} detected radio emission coincident with a relatively 
isolated filament of galaxies (in the region of the cluster 
ZwCl2341.1+0000), similar to S$_{Diff}$. Unlike S$_{Diff}$, however, the 
filament was also detected in X-rays with a L$_{X}$(0.1-2.4~keV)$\approx 
10^{44}$~erg~s$^{-1}$. Though the linear extent of S$_{Diff}$ is small 
compared to the overall size of the filament, it is in the densest region. 
The radio luminosity of S$_{Diff}$ (Table 2), assuming it is embedded in 
the z$\sim$0.04 filament, is also several orders of magnitude above the 
radio vs. X-ray luminosity correlation for known relics (Figure 
\ref{lrvslx}). Therefore, if S$_{Diff}$ is indeed caused by processes in 
the ICM or IGM, the lack of X-ray emission poses the same problem as it 
does for N$_{Diff}$. We do not have an active radio galaxy nearby, 
however, to offer an explanation for S$_{Diff}$'s increased radio 
luminosity.

\subsection{Implications} A consequence of attributing N$_{Diff}$'s 
abnormally high radio luminosity to the presence of relic radio plasma is 
that \emph{any} relic that forms in the presence of preexisting magnetized 
plasma will show an increased luminosity. A signature of this scenario 
might be in the scatter of the currently observed relics, which could be 
correlated with the availability of relic plasma from current/past AGN 
activity. However, the Coma relic 1253+275, which is a relic that is 
seemingly being fed magnetized plasma from the NAT source NGC 4789 
\citep{enss01}, is less radio luminous than other relics with similar 
X-ray luminosity (Figure \ref{lrvslx}). This is contrary to our hypothesis 
that fossil AGN plasma increases the radio luminosity of relic sources. 
There are, however, many factors that contribute to the presence and 
strength of synchrotron emission at a structure formation shock. 
\cite{mini01} found scatter of the same order as Figure \ref{lrvslx} for 
the 1.4~GHz emission of primary electrons in their simulations. This is 
after integrating the radio luminosity over a spherical volume with a 
radius of 1.3$~h^{-1}$~Mpc, and without including relic plasma from past 
AGN activity. Since \cite{mini01} used a fixed radius for all the 
clusters, the radio luminosities they found depend on the details of the 
current dynamical state and the location of shocks within (or just 
outside) each cluster. In practice, radio relics represent only a single 
(perhaps partially) illuminated shock front associated with a cluster. It 
is thus no surprise that there is such a large scatter in the 
observational correlation.

What then makes these two sources, N$_{Diff}$ and S$_{Diff}$, special? Why 
have radio relics only been found around rich X-ray clusters up until now?  
Part of the observational problem is most certainly selection and bias 
effects. Thus far most searches for relics have focused on rich X-ray 
emitting clusters, an exception being \cite{rudn06}, from which 0809+39 
was found. Even \cite{dela06} found only a handful of new candidates, but 
this could be the tip of a larger distribution that is currently below the 
NVSS and WENSS surface-brightness limits. On the theoretical side, most 
simulations that include synthetic radio observations, e.g., \cite{mini01} 
and \cite{pfro08}, do not include pre-existing relativistic plasma from 
past/current AGN. \cite{hoef04} simulated the merger of two M~$\approx 
1.6\times 10^{13}~M_{\sun}$ clusters of galaxies in order to track the 
revival of relic radio plasma. They found that efficient re-acceleration 
only occurred in regions where the ratio of magnetic pressure to gas 
pressure is P$_{B}$/P$_{gas}$ $<$ 0.01, which explains why this process is 
inefficient in the inner regions of clusters where P$_{B} \approx$ 
P$_{gas}$. One property that distinguishes N$_{Diff}$ and S$_{Diff}$ from 
other known relics is that they reside in poor environments. Perhaps poor 
clusters and groups do not confine relativistic plasma as well, so that 
radio lobe plasma from AGN can be driven farther (and faster) into the low 
density regions than they would in rich clusters (we are speaking here 
about driven jets, not buoyant forces). For adiabatic compression to be 
most effective, one wants ``younger" electrons, higher compression factors 
(steeper shocks in cooler gas), and lower magnetic fields (i.e., less 
radiative losses; \citealt{enss01}). The first two of these conditions can 
be effectively achieved in the above scenario. Simulations focusing on the 
evolution of relic AGN plasma in low-density environments are needed in 
order to fully assess the plausibility of this idea.

\section{Summary} We have presented detailed radio observations of the 
diffuse source 0809+39 in an attempt to discover the origin of the 
synchrotron emission. To summarize the key messages:\\
\noindent$\bullet$ Evidence points toward N$_{Diff}$ being a radio relic, 
i.e. shock excited synchrotron emission, related to a poor z$\sim$0.2 group of galaxies.\\
\noindent$\bullet$ S$_{Diff}$'s origin is ambiguous, though its coincidence with 
a filament of galaxies at z$\sim$0.04 makes it possible that it could be 
synchrotron emission from filamentary large-scale structure or relic 
emission from an extinct radio galaxy within the filament. \\
\noindent$\bullet$  Both N$_{Diff}$ and S$_{Diff}$ are more radio luminous than their X-ray properties would suggest, given the apparent P$_{\nu}$vs.L$_{X}$ correlation of known radio relics. \\ 
\noindent$\bullet$ Total energies in N$_{Diff}$ and S$_{Diff}$ are comparable 
to luminous, diffuse radio galaxies, and could be the result of adiabatic 
compression/reacceleration of past AGN activity. \\ 
\noindent $\bullet$ When analyzing diffuse radio emission beyond the 
environments of rich galaxy clusters, determining the true physical origin of 
these structures is non-trivial, especially at very low surface-brightness 
levels where emission related to large-scale structure is expected.

The issues presented here highlight the difficulty in finding a unified 
physical model for radio relics. Detailed observations of the radio spectrum, 
coupled with deep X-ray observations, are needed for a large sample of radio 
relics in both poor and rich environments in order to determine such a model.

\acknowledgments
   We gratefully acknowledge help and advice from G. de Bruyn and M. Brentjens 
during our RM-Synthesis analysis of the WSRT data.  We thank K. Delain for help 
in setting up the VLA and WSRT observations. Partial support for this work at 
the University of Minnesota comes from the U.S. National Science Foundation 
grants AST~0307600 and AST~0607674.

The Very Large Array is a facility of the National Science Foundation, operated 
by NRAO under contract with AUI, Inc. We also acknowledge the use of NASA's 
SkyView facility\footnote{(http://skyview.gsfc.nasa.gov)} located at NASA 
Goddard Space Flight Center. The Westerbork Synthesis Radio Telescope is 
operated by the ASTRON (Netherlands Foundation for Research in Astronomy) with 
support from the Netherlands Foundation for Scientific Research (NWO). The 
Digitized Sky Surveys were produced at the Space Telescope Science Institute 
under U.S. Government grant NAG W-2166. Archival observations obtained from 
XMM-Newton, an ESA science mission with instruments and contributions directly 
funded by ESA Member States and NASA, and ROSAT archives from HEASARC. Funding 
for the SDSS and SDSS-II has been provided by the Alfred P. Sloan Foundation, 
the Participating Institutions, the National Science Foundation, the U.S.  
Department of Energy, the National Aeronautics and Space Administration, the 
Japanese Monbukagakusho, the Max Planck Society, and the Higher Education 
Funding Council for England.  The SDSS Web Site is http://www.sdss.org/.  The 
SDSS is managed by the Astrophysical Research Consortium for the Participating 
Institutions.

\onecolumn
\appendix 

\clearpage

\clearpage

\clearpage

\clearpage

\clearpage

\clearpage

\clearpage

\clearpage

\clearpage

\clearpage

\clearpage

\clearpage

\clearpage

\clearpage

\clearpage

\clearpage

\clearpage

\begin{deluxetable}{ccccccc}
\tabletypesize{\scriptsize}
\tablecaption{Properties of the Diffuse Components of 0809+39}
\tablewidth{0pt}
\tablehead{
\colhead{Source} & \colhead{P-band flux}\tablenotemark{a} & 
\colhead{L-band flux}\tablenotemark{a} & \colhead{FWHM Size} & 
\colhead{Frac. 
Pol. 351 MHz} & \colhead{Mean $\alpha$} & \colhead{$\phi$} \\
\colhead {} & \colhead{(mJy)} & \colhead{(mJy)} & \colhead{\arcsec} & \colhead{\%} & \colhead{} & \colhead{$rad~m^{-2}$} } \startdata

N$_{Diff}$ & 178$\pm$0.7 & 37.8$\pm$0.7 & 200 & 19 & -1.12 & +12 \\
S$_{Diff}$ & 136$\pm$1.1 & 24.8$\pm$1.0 & 490 & $<9$~(3$\sigma$) & -1.23 & $\sim$ +6\tablenotemark{b} \\

\enddata
\tablenotetext{a}{Errors taken from final maps $\sigma_{rms}$ and does not include the uncertainty in the total intensity calibration}
\tablenotetext{b}{Typical local galactic value}
\end{deluxetable}

\begin{deluxetable}{ccccccc}
\tabletypesize{\scriptsize}
\tablecaption{Distance Dependent Properties of 0809+39}
\tablewidth{0pt}
\tablehead{
\colhead{Region} & \colhead{z} & \colhead{$P_{0.32}$} & \colhead{$P_{1.4}$} & \colhead{$L_{X}$~(0.1-2.4~keV)} &
\colhead{Physical Size} & \colhead{$B_{min}$} \\
\colhead{} &\colhead{} & \colhead{$10^{23}$~W~Hz$^{-1}$} & \colhead{$10^{23}$~W~Hz$^{-1}$} & \colhead{log(erg~s$^{-1}$)} &
\colhead{Mpc} & \colhead{$\mu$G}
}
\startdata
N$_{Diff}$ & 0.20 & 205  & 43.4 & $<$43.0~(3$\sigma$) & 0.66 & 0.64 \\
S$_{Diff}$ & 0.04 & 5.07 & 0.99 & $<$41.5~(3$\sigma$) & 0.39 & 0.57 \\
\enddata
\end{deluxetable}

\end{document}